\documentclass[twocolumn]{aa}
\usepackage{graphicx}
\usepackage{amsmath}
\usepackage{amssymb}
\usepackage{epsf}
\begin{document}
   \title{Cluster of galaxies around seven radio-loud QSOs at 1$<$z$<$1.6}

   \subtitle{II. $K$-band images}

   \author{S.F. S\'anchez\fnmsep\thanks{Visiting Astronomer, German-Spanish Astronomical Centre, Calar Alto, operated by the Max-Planck-Institute for Astronomy, Heidelberg, jointly with the Spanish National Commission for
Astronomy.}
          \inst{1}
          \and
          J.I. Gonz\'alez-Serrano\inst{2}
          }

   \offprints{S.F. S\'anchez}

   \institute{Astrophysikalisches Institut Potsdam (AIP)\\
		An der Sternwarte 16 \\
                D-14482 Potsdam (Germany)\\
              \email{ssanchez@aip.de}
         \and
	     Instituto de F\'\i sica de Cantabria\\
	 	Universidad de Cantabria-CSIC\\
		Avd. de Los Castros S/N\\
		 35005-Santander (Spain)\\
             \email{gserrano@ifca.unican.es}
             }

   \date{Received June --, 2002; accepted Month Day, Year}

   \abstract{
We have conducted a NIR study of the environments of seven radio-loud
quasars at redshifts 1$<$z$<$1.6. In present paper we describe deep $K$
band images obtained for the fields of $\sim$6$\times$6 arcmin around
the quasars with 3$\sigma$ limiting magnitudes of $K\sim$20.5. These
fields were previously studied using deep $B$ and $R$ band images
(S\'anchez \& Gonz\'alez-Serrano 1999). Using together optical and NIR
data, it has been found a significant excess of galaxies which
optical-NIR colours, luminosity, spatial scale, and number of galaxies
are compatible with clusters at the redshift of the quasar.

We have selected a sample of cluster candidates analyzing the $R-K$
vs. $K$ diagram. A $\sim$25\% of the candidates present red
optical-NIR colours and an ultraviolet excess. This population has
been also found in clusters around quasars at the same redshifts
(Tanaka et al. 2000; Haines et al. 2001). These galaxies seem to
follow a mixed evolution: a main passive evolution plus late
starformation processes. The quasars do not inhabit the core of the
clusters, being found in the outer regions.  This result agrees with
the hypothesis that the origin/feeding mechanism of the nuclear
activity were merging processes. The quasars inhabit the
region were a collision is most probably to produce a merger.

   }

   \maketitle

%

\section{Introduction}

There is much evidence for a connection between the QSO activity
and clustering of galaxies (Stockton 1982; Yee \& Green 1984, 1987;
Gehren et al. 1984; Hutchings et al. 1984; Hintzen 1984; Yee 1987;
Ellingson et al. 1991; Hintzen et al. 1991; Yee\& Ellingson 1993;
Fisher et al. 1996; Yamada et al. 1997; Hall et al. 1998; Hall \&
Green 1998; S\'anchez \& Gonz\'alez-Serrano 1999; Cimatti et al. 2000;
Wold et al. 2000). It seems that interaction/merging processes amoung
cluster galaxies could rule a fundamental role in the origin/feeding
process of the nuclear activity. The large fraction of host galaxies
of radio quasars with distorted morphologies (e.g. Disney et al. 1995;
Hutchings \& Neff 1997; S\'anchez 2001; S\'anchez et al. 2002), their
large luminosities at $z\sim$1 (Carballo et al. 1998; S\'anchez et
al. 2002, and references therein), direct evidences in their spectra
(Nolan et al. 2001), and the velocity distribution of galaxies around
QSOs (Heckman et al. 1984; Ellingson et al. 1991) give support to
this hypothesis (Yee \& Ellingson1993).

Yee \& Green (1987), Ellingson et al. (1991), Yee \& Ellingson et
al. (1993) and Wold et al. (2000) have shown evidences of the
connection between radio emission and galaxy clustering around QSOs.
They found differences between the environment of radio-loud and
radio-quiet quasars. Since radio-loud QSOs are found in clusters of
galaxies with an Abell richness class between 0-2 (Abell 1958),
radio-quiet relatives inhabit groups of galaxies of richness similar
or even lower than 0. However, these results were based in studies of
overdensities of galaxies around low-$z$ quasars ($z<$0.7). At higher
redshifts, the number of studies decreases (Hitzen et al. 1991; Boyle
\& Couch 1993; Hall \& Green 1998), or were based in a reduced number
of objects (Hutchings et al. 1993; Yamada et al. 1997; S\'anchez \&
Gonz\'alez-Serrano 1999; Teplitz et al. 1999; Cimatti et
al. 2000). However, the differences between the environment of
radio-loud and radio-quiet quasars appear to persist at high-$z$
(1$<z<$2).

Hall et al. (1998) and Hall \& Green (1998) showed that the excess of
galaxies around their sample of 31 radio-loud QSOs at 1$<z<$2 seems to
present a two component distribution, with a peak around the quasars
at $r<$35$\arcsec$ and a smooth component that extends to r$\sim$100$\arcsec$ (his
maximun sampled distance). They explained this two-component
distribution as a cluster of galaxies embebbed in a large scale
structure. In S\'anchez \& Gonz\'alez-Serrano (2001), hereafter Paper
I, we have analyzed the deep $B$ and $R$ band images of seven radio
loud quasars at 1$<$z$<$1.6, with a larger sampled spatial region,
finding a similar result. However, the explanation to this peculiar
distribution was found more simple, since the quasars seem not to be
in the center of the overdensities of galaxies. Once corrected for the
correct overdensity centers, the two-component distribution blurs, and
a typical cluster overdensity profile remains.

We present here deep $K$-band images of fields of $\sim$6$\arcmin$$\times$6$\arcmin$
around the seven radio quasars studied in Paper I (Section 2). We have
looked for possible overdensities of galaxies also in this band
(Section 3), and, together with the $B$ and $R$ data, we have selected
a sample of candidates to cluster members (Section 4). We have tried
to find if the results presented in Paper I could be confirmed with
these new data. The combined $B$, $R$ and $K$ band data for the
different galaxies detected have been used to study the
characteristics of these cluster candidates (Section 5). Throughout
this article we have assumed a standard cosmology with H$_o$=50 km
s$^{-1}$ Mpc$^{-1}$ and q$_o$=0.5. It is interesting to note here that
at the redshift range considered in this article the differences in
the scales were small ($<$10\%) if we use a cosmology with H$_o$=70
kms$^{-1}$ Mpc$^{-1}$ $\Omega_m$=0.3 and $\Omega_\lambda$=0.7.

\section{Observations, reduction and photometry}

Deep NIR images of the 7 quasars studied here have been obtained in
the night February the 3th 1999 at the 3.5m telescope of the Centro
Astron\'omico Hispano-Alem\'an of Calar Alto. The data were obtained
using {\it Omega Prime}, with a pixel scale of 0.396$\arcsec$/pixel,
and a field-of-view of $\sim$6'$\times$6'.  { A standard NIR
observing procedure was used: 25 unregistered exposures of 4 seconds
each were taken and the average image was registered. A number of 9
average images were obtained following a grid pattern around the
object central position, with an offset of 35 arcseconds. } Both the
target and the photometric standards were observed using the same
procedure. The whole procedure was repeated three times, which allowed
a total exposure time of 2700s.  {\it Omega} is an array of
1024$\times$1024 pixels, formed by 4 arrays of 512$\times$512
pixels. In order to avoid the posibility that the QSO image lies in
the junction of the different arrays, we have off-centered the image
30$\arcsec$ North-West.

The data were reduced using standard IRAF packages. First, a dark
frame, obtained before each sequence of exposures along the grid, was
subtracted to each registered image. Domeflats were obtained as
suggested by the {\it Omega} (and {\it Magic}) observer's manual.  The
sky-flux frame was then obtained for each sequence of 9 images,
averaging them. The flux from astronomical sources was masked by a
1$\sigma$ clipping rejection over the mean value. The sky-flux frame
was subtracted to each of the images of the sequence. Once sky
subtracted, the images of each sequence were recentred and
co-added. When more than one sequence was obtained for a certain
object, we recentred and coadded all the sequences to obtain the
final image.

All the images were obtained in good photometric conditions. Flux
calibration for each night was carried out using UKIRT faint standard
stars observed along the night. The error of the photometric calibration
was about $\sim$0.09 mag, the mean 3$\sigma$ limiting surface brightness 
was 21.0 mag/arcsec$^{2}$, and the mean seeing was 1.3$\arcsec$.

\section{Analysis of the data}

\subsection{Detection of galaxies}

Searching galaxies over the images were made using the { SExtractor}
program (Bert\'\i n \& Arnauts 1996). This program detects and split
objects up to a certain flux limit over the local sky background (ie,
in each object environment), producing a catalogue of objects with
their main properties. This catalogue includes the position,
ellipticity, effective surface brightness, magnitude and a
classification parameter (star/galaxy), for each detected object. The
objects were classified by a neural network which takes as input
parameters 8 areas determined by the same number of isophotes, similar
for every object, the intensity peak and the FWHM, together with the
{\it seeing }, defined as the mean FHWM of the stellar images. This
neural network determines the so-called stellar index, which is, by
definition, 0 for the galaxies and 1 for the stars. In practice a
value of 0.7 works fine to split galaxies from stars.

The detection area and significance detection limit were fixed to be
similar to values used in Paper I, being 1.4 arcsec$^2$ and
4.5$\sigma$, respectively. We have used a different number of
connected pixels and detection limit per pixel in present study due to
the differences in pixel size in optical and NIR images. The detection
process was done looking for 9 connected pixels with a flux per pixel
up to 1.5$\sigma$. Once the objects near the border of the images,
near clear foreground galaxies and dead regions of the detector were
masked, we got the final catalogues. The mean magnitude of the
faintest detected objects field by field were $\langle
K\rangle$=20.3$\pm$0.3 ($\langle B\rangle$=25.6$\pm$0.4 and $\langle
R\rangle$=24.8$\pm$0.3, for the optical images).

The parameter to determine the depth of a certain image is the
3$\sigma$ limiting magnitude for the detection area. This magnitude
can be estimated determining the standard deviation of the sky
background flux in the detection area, which yields
$K_{lim}^{3\sigma}$=20.5$\pm$0.12 for the studied images. This limit
implies a 4.5$\sigma$ detection limit of $\sim$20.1 mag, consistent
with the mean magnitude of the faintest detected objects. This a good
test of the consistency of the detection method.

This limiting magnitude is similar to the presented in other studies
of cluster of galaxies at similar redshifts and wavelenghts. E.g.,
Tanaka et al. (2000) presented a study of the overdensities of
galaxies around the radio quasar B2 1335+28 at $z$=1.1, finding a
possible cluster of galaxies at the redshift of the object. The
faintest detected galaxies in the $K$ band on that study have $\sim$20
mag.

\begin{table}
\centering
\caption{Number of field galaxies and $\gamma$ factor per circular area of  1.07arcmin$^2$ (r$\le$35$\arcsec$), for different magnitude ranges}
\label{tab:nnir}
\vspace{0.2cm}
\begin{tabular}{lcc}
\hline
\hline
Mag. Range & N$_{exp}$ & $\gamma$  \\
\hline
$K$ all &5.41$\pm$2.33 & 1.00$\pm$0.15 \\
$K<$19  &3.62$\pm$1.95 & 1.03$\pm$0.14 \\
$K<$18  &1.80$\pm$1.41 & 1.05$\pm$0.12 \\
$K<$17  &0.60$\pm$0.68 & 0.88$\pm$0.07 \\
$K<$16  &0.17$\pm$0.38 & 0.92$\pm$0.06 \\
$K<$15  &0.03$\pm$0.18 & 0.99$\pm$0.04 \\
\hline
\end{tabular}
\end{table}

\subsection{Number counts}

An appropiate determination of the number of field galaxies and its
standard deviation is fundamental for a correct detection of excess of
galaxies and its significance. As discussed in Paper I, galaxies tend
to cluster which makes the standard deviation of number counts to
depend not only in this number (like in a poisson distribution,
$\sigma$=$\sqrt{n}$). Therefore, the standard deviation depends on the
number of galaxies, $N$, and the angular scale, $r$, which can be
expressed as $\sigma (r,N)$=$\gamma (r) \sqrt{N}$. The factor $\gamma
(r)$ reflects the two point correlation and depends on the angular
scale, since it is expected that there were no angular correlation at
all scales. This factor tends to 1 at large scales (Yee \& Green
1987), which corresponds to the poisson distribution.

Both the number of field galaxies or number counts and the standard
deviation were determined directly on the images, comparing latter
with similar results from the literature. We have counted the number
of galaxies in a grid of 25 not overlapping circular areas of r=35
arcsec ($\sim$1.1arcmin$^2$). This angular scale corresponds to
$\sim$0.5Mpc at the mean quasar redshift. To minimize possible
contaminations from cluster members, we have used only the
measurements of areas at more than 70 arcsec ($\sim$1Mpc) from the
quasars. With this method we got 16 measurements of the number of
galaxies for each field, and a total of 112 measurements, which yields
a good estimation of both the number of galaxies and the factor
$\gamma$. As we explained in Paper I, this method does not guarantee
that there were no contaminations from putative clusters. However, any
possible contamination would produce an increase of both the estimated
number of field galaxies and the factor $\gamma$ and, therefore, they
would reduce the significance of the possible excesses.

\begin{figure}
\epsfxsize=9cm
\begin{minipage}{\epsfxsize}{\epsffile{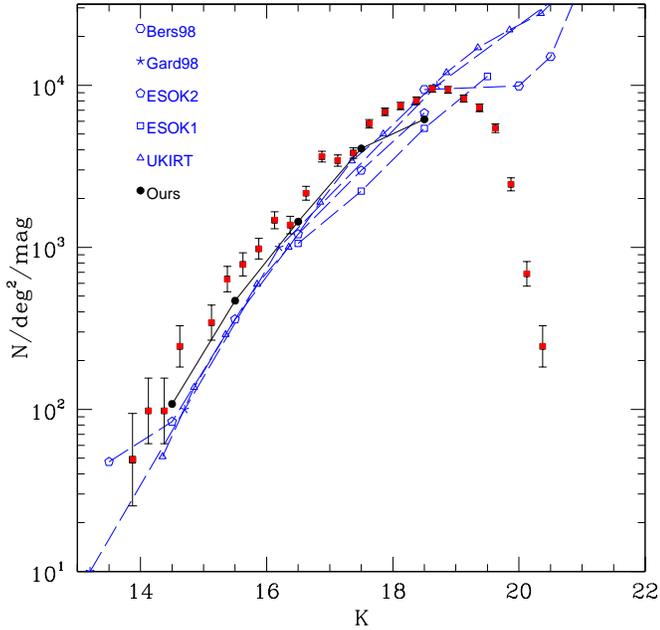}}
\end{minipage}
\caption[]{{\small
Number counts of field galaxies for different $K$-band ranges (black circles).
Results from different authors have been also presented (blue):
{\it Bers98} (Bershady et al. 1998), {\it Gard98} (Gardner
1998), {\it ESO K1,K2} (Sarocco et al. 1997) and {\it UKIRT} (Hall et
al. 1998).  Red squares show the total number of galaxies detected in 
our images, scaled to one square degree area. It is possible to see
a clear excess of galaxies for any estimation of number counts.
}}
\label{fig:n_counts_k}
\end{figure}

Table \ref{tab:nnir} lists the estimated number of field galaxies {
per circular area of 1.07 arcmin$^2$ } at different magnitude ranges
(N$_{exp}$), together with the factor $\gamma$. This value, near to
$\sim$1, is similar to the value found on the optical images
($\sim$1.08, Paper I). Figure \ref{fig:n_counts_k} shows the
distribution of number counts of field galaxies for different $K$-band
ranges (solid circles), together with different estimations extracted
from litetature. We have also plotted the total number of detected
galaxies in the different fields, scaled to one square degree area
(solid circles). It is seen that there is a clear excess of galaxies
down to $K\la$19 mag, using both our estimated number counts and
values from the literature. This excess will be discussed below. The
slope of this distribution is $\sim$0.45 in the range of magnitudes
14.5$\le$K$\le$18.5 (similar to values found in Paper I for $R$ and
$B$ band, $\sim$0.37 and $\sim$0.49, respectively). The 100\%
completeness magnitude, defined as the magnitude where the
distribution deviates from the power law, was $K\sim$19 mag. This
magnitude corresponds to objects detected with a signal-to-noise ratio
of $\sim$30$\sigma$, i.e., with photometric errors down to $\sim$0.03
mag. Both the number of field galaxies, its distribution along the
magnitude range and its slope agree with previous results published in
literature (Sarocco et al. 1997: ESO K1 and K2; Gardner et al. 1998;
Bershady et al. 1998; Hall et al. 1998:UKIRT). However, there are
small differences in the estimated number counts from different
authors, as it is seen in Figure \ref{fig:n_counts_k}. These
differences are mainly due to different image quality (different
seeing, photometric conditions... etc), small differences in the
transmission of the filters, and the use of different method for the
detection and classification of the objects (Hintzen et al. 1991; Kron
1980). These are the main reasons to use number counts estimated from
the same images where the excess of galaxies would be studied.

\subsection{Excess of galaxies}

As we quoted above, in Paper I we found an excess of galaxies around
the radio quasars studied also in the present article using deep $B$
and $R$ images. This excess, significant for both bands, had a clear
dependence with the magnitude range, being stronger for the faintest
range of magnitudes (up to 3$\sigma$ for $R>$22 and $B>$22.5), which
induced us to think that it was due to galaxies at the redshift of the
quasar. The peak of the excess was not found nearby the QSOs, but at a
certain distance ($\sim$70$\arcsec$, $\sim$1Mpc). In order to
confirm/reinforce this result we have repeated the analysis over the
new $K$-band images.

We have measured the number of galaxies in two different angular
scales around the quasars, all the image (which correspond to an
angular scale of $r_1$$\la$170$\arcsec$), and nearby the QSOs,
$r_2\la$35$\arcsec$, and at different magnitude ranges. The number of
galaxies found around each quasar was coadded and them it was compared
with the expected number of galaxies. It was then splitted in groups
by their ranges of magnitudes, down to the 100\% completeness
magnitude, and an additional group for the galaxies faintest than this
magnitude. It is important to note here that all the galaxies, even
the galaxies fainter than this completeness magnitude, were detected
with a signal-to-noise up to 4.5$\sigma$. Moreover, the number of
expected galaxies were estimated from the same images, using the same
detection/classification method, and therefore it was affected by
the same incompleteness effects.

Table \ref{tab:excess} lists the number of galaxies detected for both
angular scales ($r<$35$\arcsec$ and $r<$170$\arcsec$), the number of
expected galaxies and significance of the excess, for each magnitude
range and for all ranges (indicated as {\it tot}). It is seen that
there is a significant excess of galaxies for the largest scale, and
mainly for the range of magnitudes $K$$<$19. The excess is less
significant for the smallest scale, nearby the QSO. This is expected
from results of Paper I, where we found that the peak of the excess
was not at the QSO position. We will discuss this topic below.

Using present results and results from Paper I together, it seems that
the excess is not due to foreground galaxies, since there is no
significant excess of galaxies in the brightest ranges for the optical
magnitudes, $R<$21 and $B<$22.5 (e.g., Ben\'\i tez et al. 1995). The
excess is most probably due to galaxies at the redshift of the
quasars: (i) the mean optical-NIR colours of the excess galaxies are
$R-K$$\sim$4.5-6.5, assuming that they are the same galaxies in both
bands, and (ii) at the mean redshift of the quasars ($z\sim$1.4),
galaxies with magnitudes $B\sim$22.5, $R\sim$22.5 and $K\sim$18 would
have $M_B\sim$--18.6, $M_R\sim$--20.9 and $M_K\sim$--24.79, which
are not unusual values (e.g., Bromley et al. 1998; Muriel et al. 1998;
Mobasher et al. 1993). We have used typical $k$-correction for
elliptical galaxies, using the $c$ model (Bruzual 1983), with the
GISSEL code (Bruzual \& Charlot 1993), assuming a formation redshift
of $z_{\rm for}$=5, and only one starburst period of 1Gyr, and a Salpeter
initial mass function (Salpeter 1955), with a mass range between 0.1
and 125 M$_{\odot}$. Furthermore, Hall \& Green (1998) concluded that
galaxies with $r>$23 and colours $r-K>$5 could be associated with
quasars with a $z\sim$1.5. Taking in account the proper conversion
between $R$ and $r$ bands, these galaxies would have $R>$22.3 and
$R-K>$4.3, values similar to those presented here.

   \begin{table*}
      \caption[]{Number of galaxies at $r_1$$<$170$\arcsec$ and $r_2$$<$35$\arcsec$ around the quasars for all the fields}
         \label{tab:excess}

      \[
         \begin{tabular}{lrrrrrrr}
            \hline
            \noalign{\smallskip}
            \hline
            \noalign{\smallskip}
{Rango de mag.}
&{N$_{\rm tot}$ }
&{N$_{\rm esp}$ }
&{$n\sigma_{\gamma}^{a}$}
&
&{N$_{\rm tot}$ }
&{N$_{\rm esp}$ }
&{$n\sigma_{\gamma}^{a}$}\\

            \hline
            \noalign{\smallskip}

 & &$r$$<$170$\arcsec$& & & &$r$$<$35$\arcsec$& \\
\cline{2-4}
\cline{6-8}
$K${\scriptsize tot}&1076& 893.42&6.11& & 51& 37.87&2.14\\
\hline
$K>$19&305&295.60 &0.55& &14&12.53&0.42 \\
$18.0<K\le$19.0&401&300.56&5.79& &14&12.74&0.35 \\
$17.0<K\le$18.0&227&204.78&1.55& &14& 8.68&1.80 \\
$16.0<K\le$17.0&107& 66.06&5.03& & 6& 2.80&1.92 \\
$K\le$16.0     & 36& 26.42&1.86& & 3& 1.12&1.77 \\
            \noalign{\smallskip}
            \hline
         \end{tabular}
      \]
\ \ \ $^{a}$ $\gamma=$1.0 

\end{table*}

In Paper I we found that, once corrected for the off-centering, the
angular scale of the overdensity has a peak of $\sim$30-40$\arcsec$
and a smooth excess until 160$\arcsec$-170$\arcsec$. If it is due to
galaxies at the redshift of the quasars, the peak would be at
$\sim$350 kpc, extending down to $\sim$1500 kpc. These scales were
compatible with physical scales of well known clusters at lower
redshifts (e.g. Dressler 1980; Ellingson et al. 1991). The overdensity
is due to $\sim$20-40 galaxies for each field, which would correspond
to not very dense clusters, with an Abell richness between 0 and 1
(Abell 1958). In summary, magnitudes, colours, physical scales and
number of galaxies are compatible with clusters of galaxies at the
redshift of the QSOs.

\section{Colours of the excess galaxies}

\subsection{Distribution of galaxies in ranges of colours}

The overdensity found in previous section and Paper I was mainly due
to galaxies with magnitudes $B$$>$22.5, $R$$>$22 and
$K$$\sim$16-18. If these galaxies are the same for all the bands,
their mean colours were $B-R\gtrsim$0.5 and $R-K$$\sim$4-6. The $R-K$
colours are compatible with a population of old galaxies, similar to
well known early-type galaxies that dominate cluster population, at
least down to $z<$1.3 (Visvanathan \& Sandage 1977; Bower et al. 1992;
Arag\'on-Salamanca et al. 1993; Kodama \& Arimoto 1997). However, even
in the case that the $B-R$ colours are considered a lower limit (since
the number of detected galaxies was lower in $B$ than in $R$, Paper
I), we have to admit that a fraction of these galaxies deviate from
pasive evolution, formed at lower redshifts or/and present a
starformation process at recent epochs.

In order to determine the range of colours of the galaxies that
produces the excess, we have made an analysis similar to the previous
one: (i) estimate the number of expected galaxies at different colour
ranges, and (ii) look for overdensities at $r$$<$170$\arcsec$ from the
QSO, for these colour ranges. It is needed to cross-correlate the
different catalogues of galaxies for each band, to produce a unique
catalogue that includes both the $B$, $R$ and $K$ band
magnitudes. This would reduce the sampled region to
5$\arcmin$$\times$5.4$\arcmin$, due to the above mentioned
off-centering of the QSOs in the NIR band images.

Figure \ref{fig:col_ex1} shows the distribution of relative excess of
galaxies for the different colour ranges. The range of expected
colours for old elliptical galaxies ($z_{\rm for}\gtrsim$4.5) at the
redshift of the QSOs has been represented by a shaded vertical
band. It is seen that the excesses present at least a peak within the
range of these colours, and for 5 of the 7 fields, the excess is only
found within this region. The observed defects could be due to
differences in the mean colour distribution, field-to-field, combined
with the different depth, band-to-band. However, these figures are
a good estimate of the distribution of excesses at different ranges
of colours. These results show that the excesses were due to galaxies
which optical-NIR colours are compatible with being old galaxies at
the redshift of the quasars. A similar analysis was used by Tanaka et
al. (2000) to determine the existence of a cluster around a quasar at
$z\sim$1.1. Although spectroscopical confirmation is needed, our data
indicate a similar result.

For 2 of the 7 fields the excess extends to bluer $R-K$ values than
expected. Excesses partially due to galaxies at redshifts lower than
the quasar redshift could explain this result. Another possibility
could be the presence of blue galaxies in the cluster around the
quasar. Tanaka et al. (2000) studied the objects around a quasar at
$z\sim$1.1, using broad-band $RIK$ images together with narrow band
images ([OII]$\lambda$3727\AA, centered at the quasar redshift), from
Hutchings et al. (1993), finding a cluster at the quasar
redshift. They found that the $R-K$ distribution was wider than
expected for a sample of old elliptical galaxies, indicating a recent
star formation, possible due to a collision process. Butcher \& Oemler
(1978,1984) observed that the optical colours of cluster galaxies tend
to bluer with the redshift. This effect, known as the Butcher-Oemler
effect, has been found in a large number of clusters at low
redshifts. Our results, like Tanaka et al. (2000), could be the
extension of this effect to beyond $z>$1.

\begin{figure*}
\begin{center}
\epsfxsize=8cm
\begin{minipage}{\epsfxsize}{\epsffile{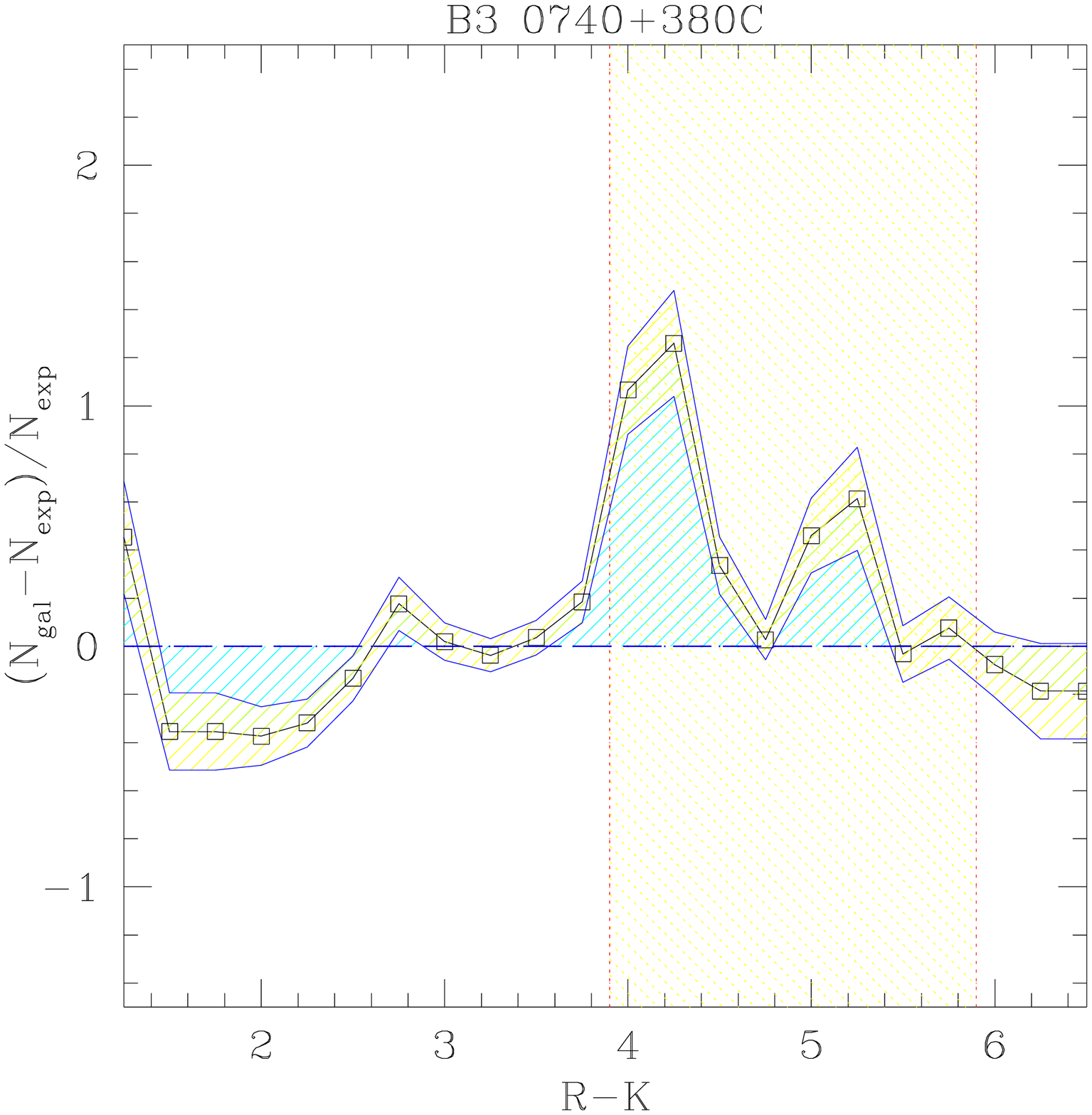}}
\end{minipage}
\epsfxsize=8cm
\begin{minipage}{\epsfxsize}{\epsffile{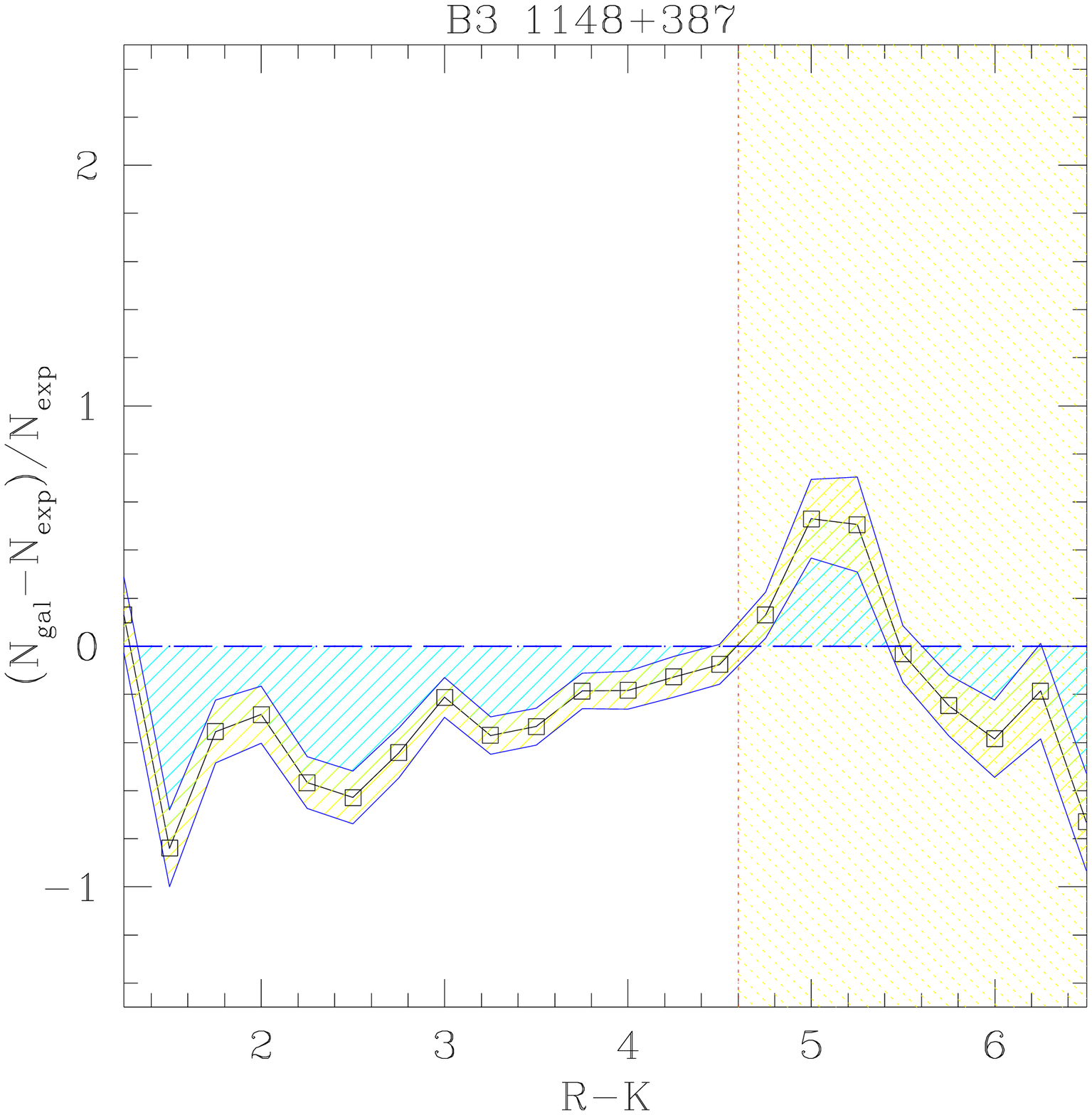}}
\end{minipage}
\epsfxsize=8cm
\begin{minipage}{\epsfxsize}{\epsffile{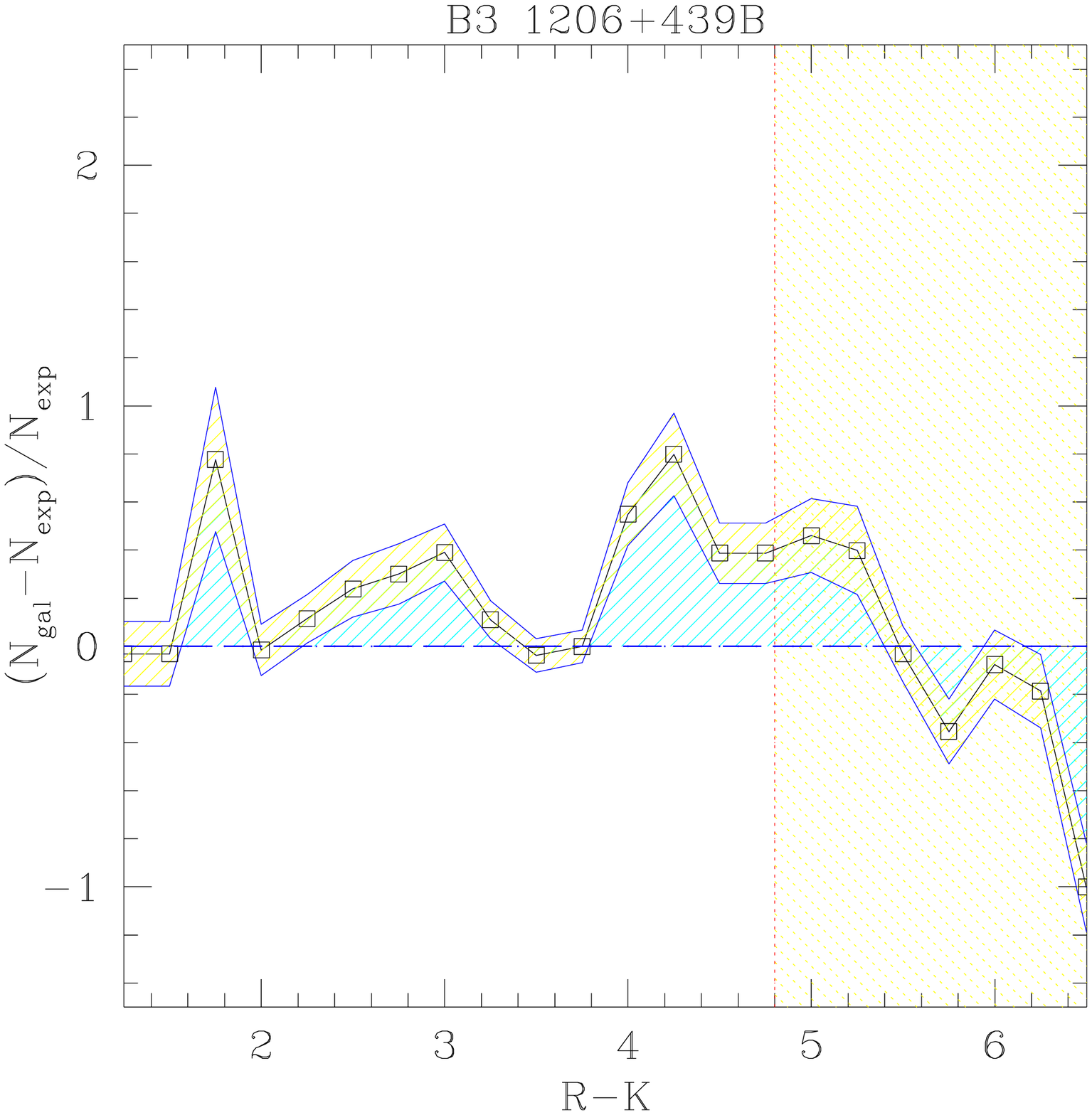}}
\end{minipage}
\epsfxsize=8cm
\begin{minipage}{\epsfxsize}{\epsffile{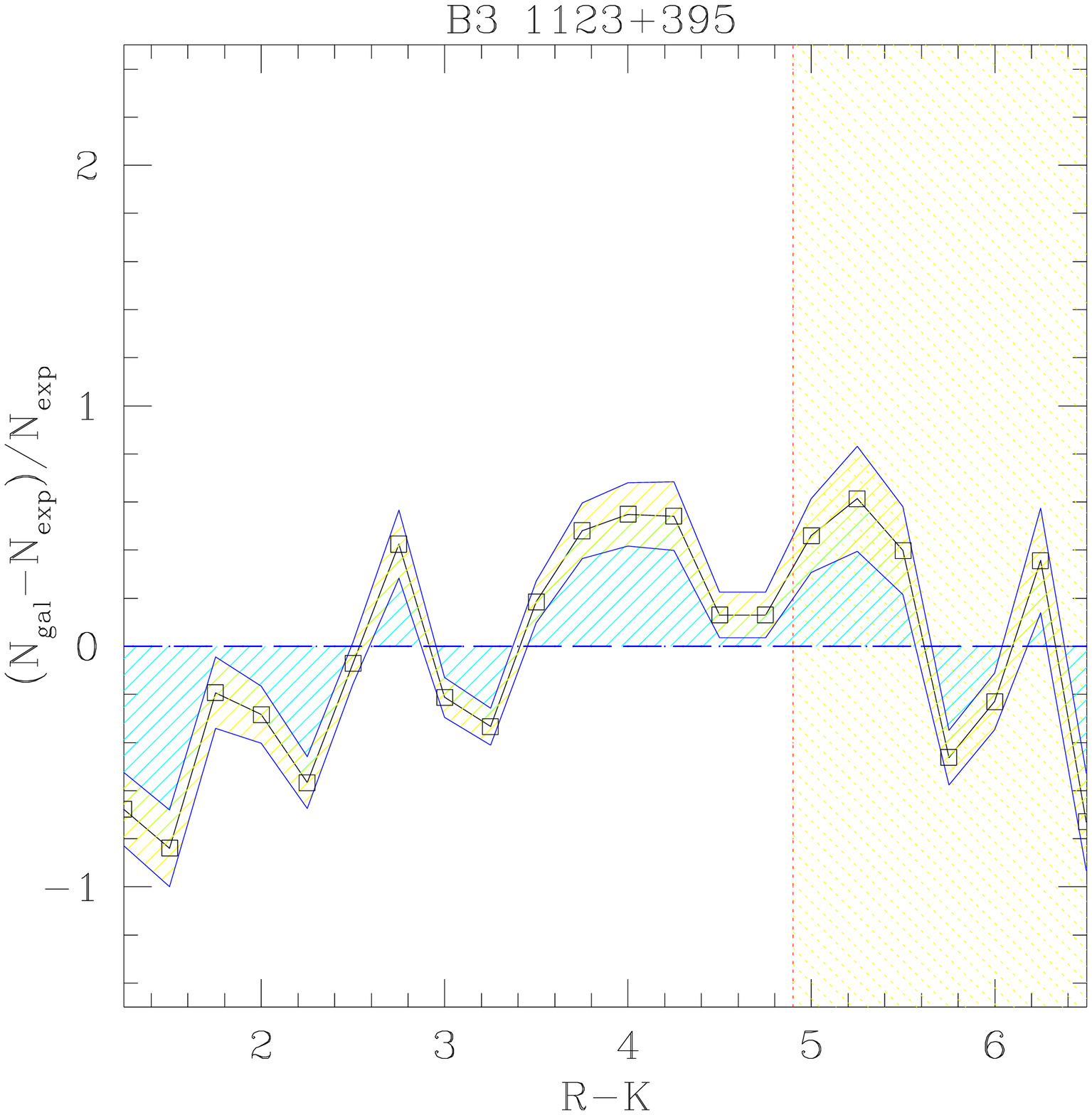}}
\end{minipage}
\end{center}
\caption[]{{\small
Galaxies excess distribution at $r$$<$170$\arcsec$ from the QSOs in
ranges of $R-K$ colours. The vertical yellow shaded region
corresponds to the range of colours expected for old elliptical
galaxies at the redshift of the QSOs.}}

\label{fig:col_ex1}
\end{figure*}

\addtocounter{figure}{-1}
\begin{figure*}
\begin{center}
\epsfxsize=8cm
\begin{minipage}{\epsfxsize}{\epsffile{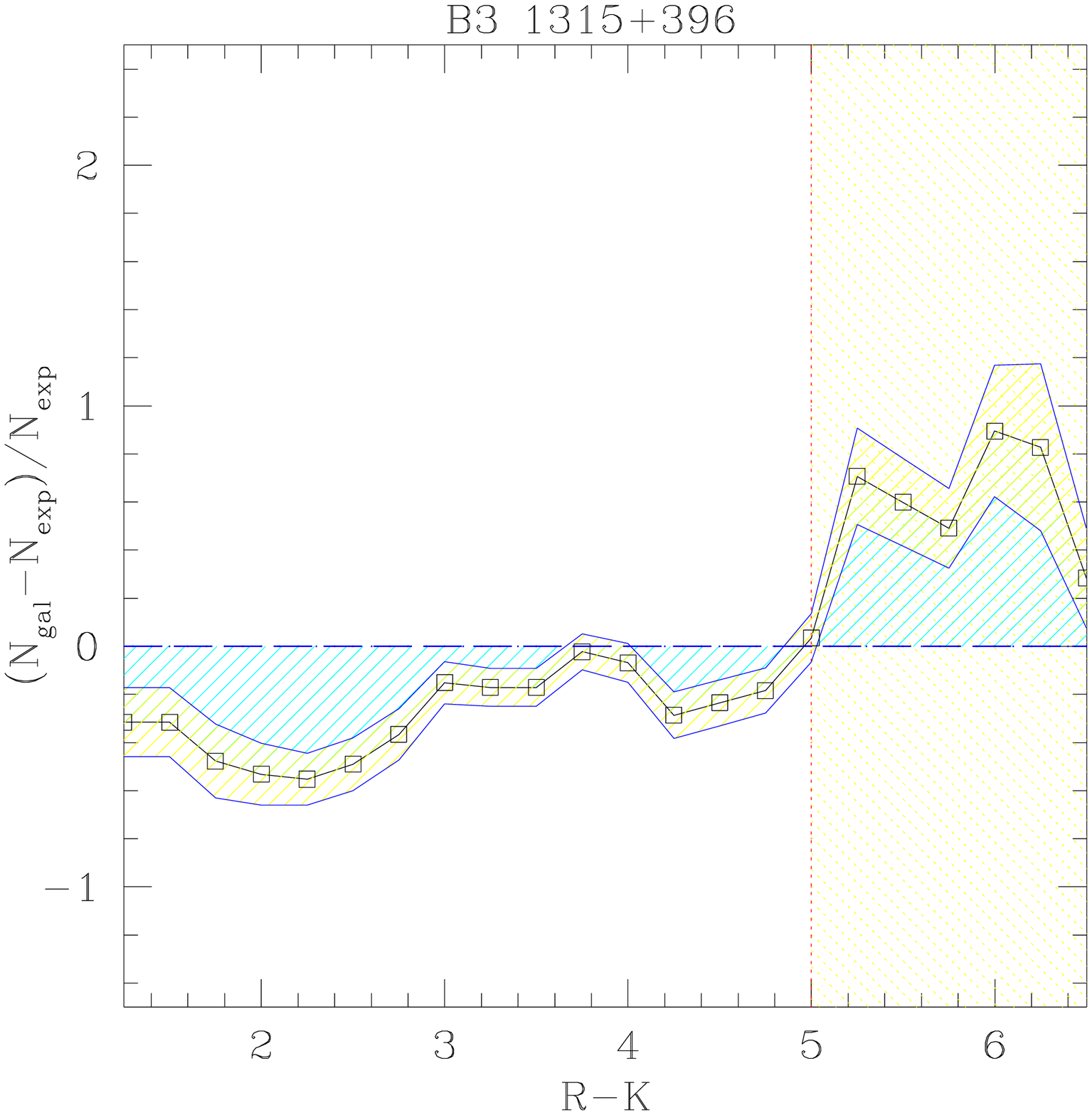}}
\end{minipage}
\epsfxsize=8cm
\begin{minipage}{\epsfxsize}{\epsffile{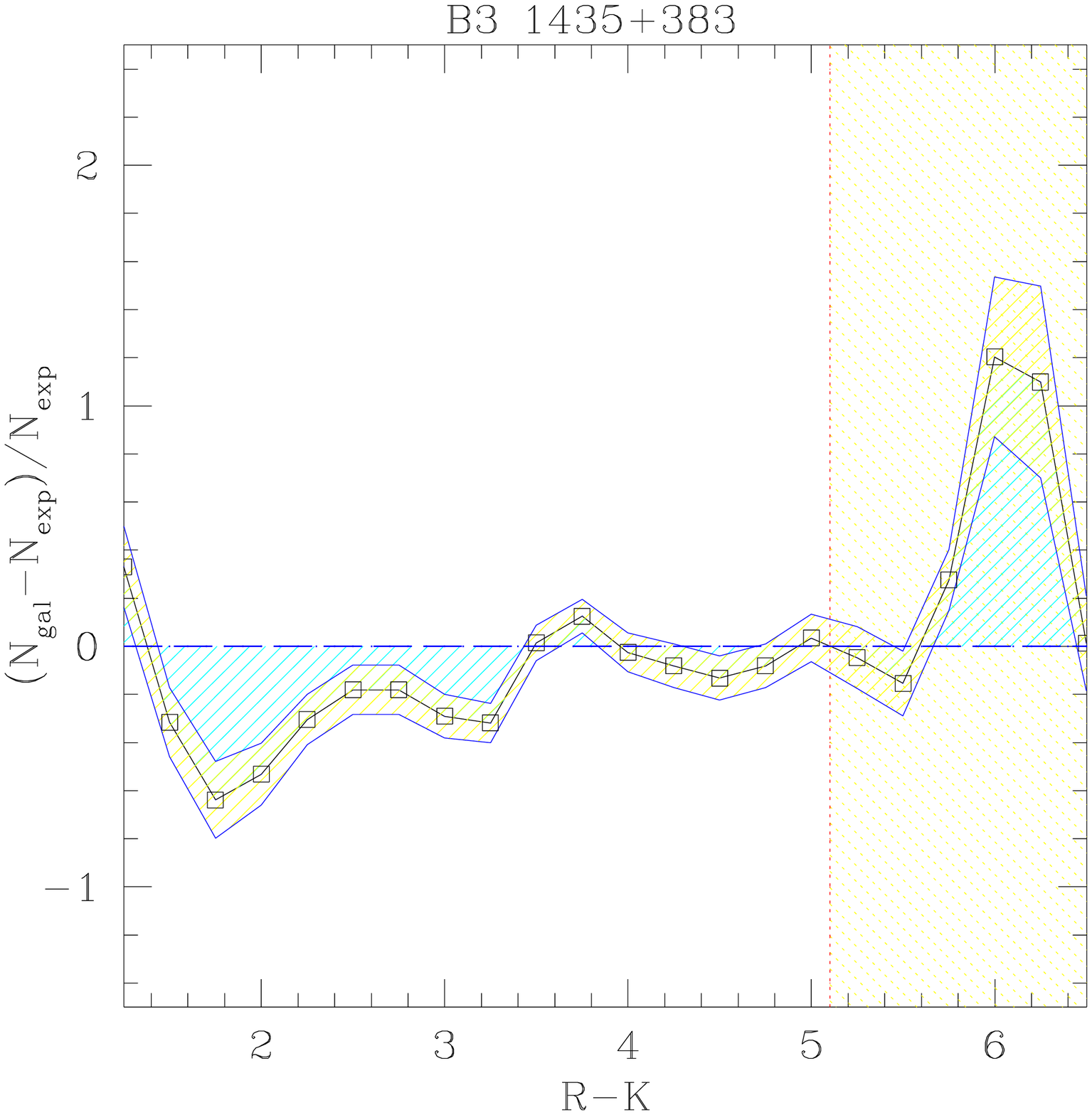}}
\end{minipage}
\epsfxsize=8cm
\begin{minipage}{\epsfxsize}{\epsffile{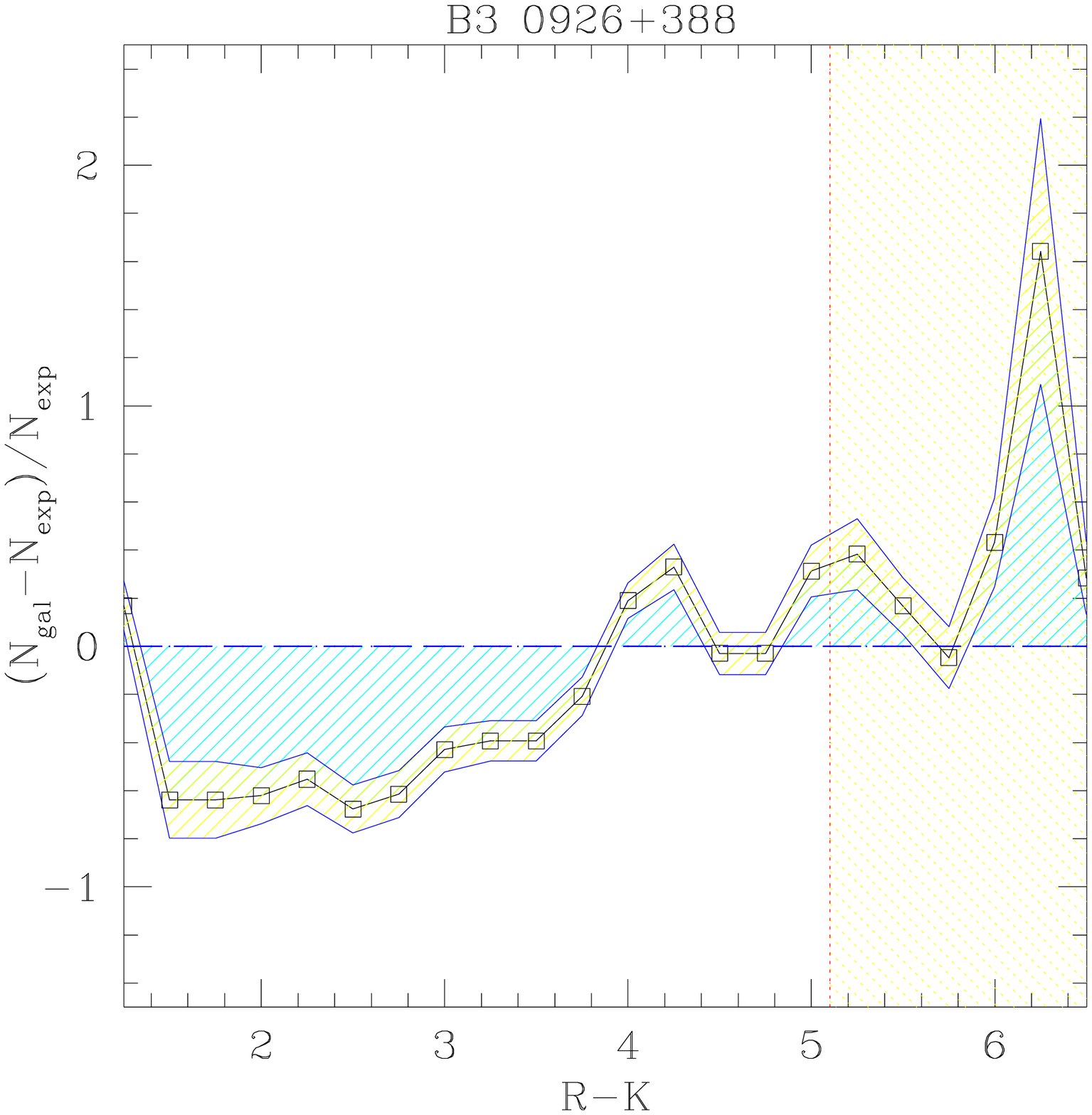}}
\end{minipage}
\end{center}

\caption[]{{\small Continued}}

\label{fig:col_ex2}
\end{figure*}

\subsection{Colour-Magnitude relation}

It is well known that there is a relation between the optical-NIR
colours and the NIR magnitudes for the cluster galaxies, compatible
with a population of old galaxies (e.g., Stanford et al. 1998; Kodoma
et al. 1998). The slope of this relation is constant along a wide
range of redshifts, 0.02$<z<$1.27, and there is evidence that it
remains constant at higher redshifts (L\'opez-Cruz 1997; Kodoma et
al. 1998). This limits significantly the formation redshift of the
bulk of stars up to $z_{\rm for}$$>$2.5. It is necessary to note here
that recent studies analyzed more exotic models, where the cluster
members do not form at the same redshift, but this $z_{\rm for}$
depends on the the galaxy luminosity (Ferreras et al. 1999; Ferreras
\& Silk 2000). This process could produce a conspiracy in which the
observational results were the same. We do not attempt to determine
the physic nature of this relation in present study, but it is
necessary to note here that any model predict the existence of this
relation.

The best situation to determine if the excess galaxies constitute a
cluster would be to know the {\it redshift}, spectroscopic or at least
photometric one. We could not afford an adequate photometric redshift
estimation with the present data. Instead, we used the CM relation to
select a sample of cluster candidates, and study their
properties. This procedure have an inherent misclassification,
including in cluster candidates galaxies that do not belong to the
cluster, and excluding cluster members. Taking into account results
from Figure \ref{fig:col_ex1}, we extimated that the fraction of
missclassified objects would be lower than $\la$20\%. This fraction
could be larger for the fields around B3 1206+439B and B3 1123+395, in
case of a the presence of a population of blue galaxies, similar to
found by Tanaka et al. (2000). To select only the red galaxies (around
the CM relation), could be considered as a conservative selection.

It would be interesting to test if it is possible to resconstruct the
CM relation for our clusters, before making the proposed selection. We
have followed the procedure proposed by Tanaka et al. (2000), used to
reconstruct the CM relation for a cluster at $z\sim$1.1. This
procedure was based on the assumption that the CM correlation were due
to bulk formation of the cluster population at a certain redshift (at
least $z_{\rm for}\gtrsim$2.5), and its slope was due to the relation
between the galaxy metallicity and luminosity/mass (Kodoma \& Arimoto
1997; Ferreras \& Silk 2000). Therefore, the galaxies that produce
this relation should be old, red in both the optical and optical-NIR
colours, and faint in the optical bands.

\begin{figure*}
\begin{center}
\epsfxsize=16cm
\begin{minipage}{\epsfxsize}{\epsffile{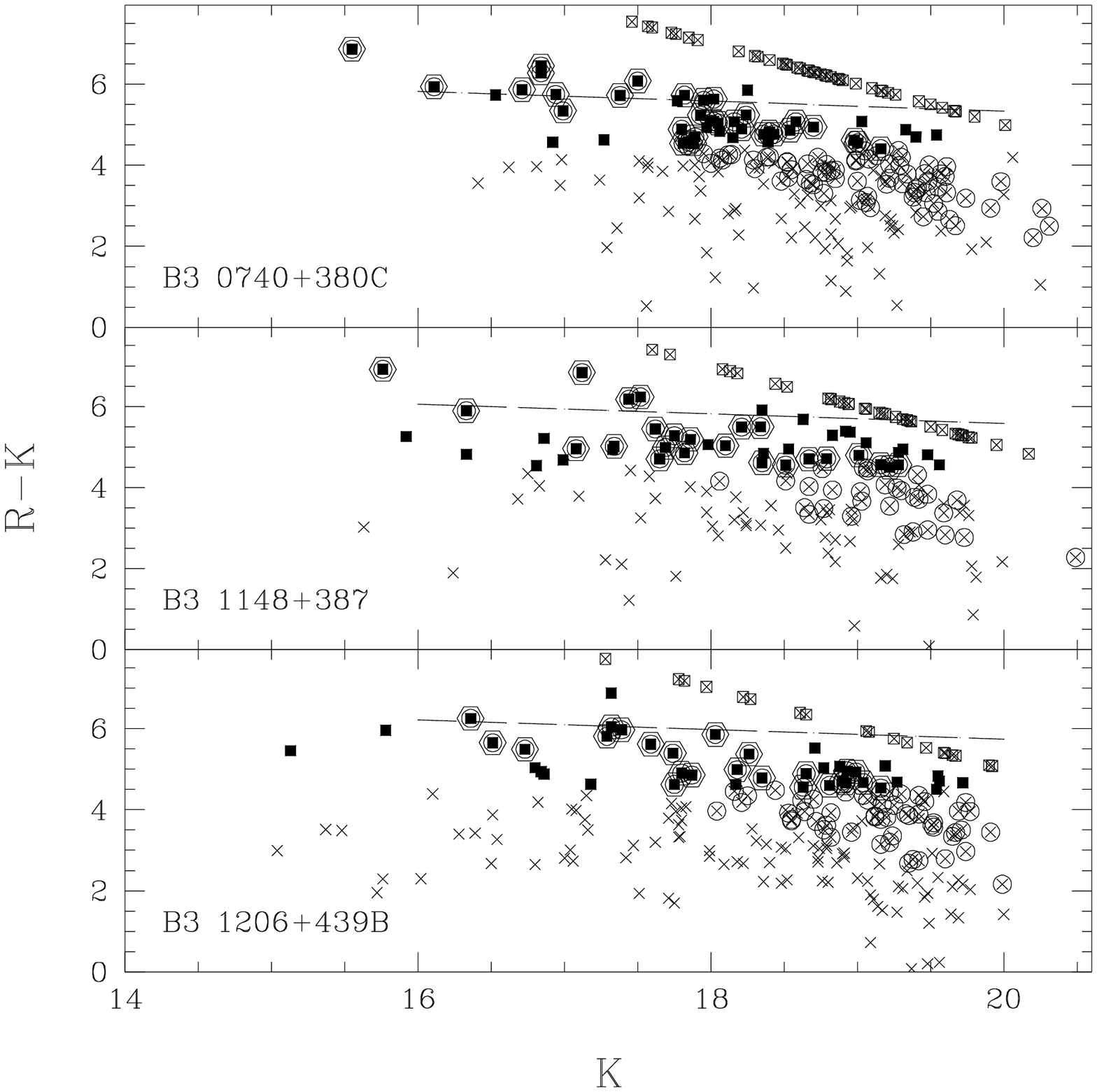}}
\end{minipage}
\caption[]{{\small
$R-K$ colour distribution versus de $K$ magnitudes for the
galaxies detected around the quasars. Red open squares show the lower limit
of the $R-K$ color for the galaxies detected in the $K$ but not in the 
$R$ band. Solid blue squares show the galaxies candidate to cluster members.
Green circles show the faint galaxies in the optical range, and, among them
the selected as candidates were indicated with open hexagons. 
Discontinous line shows the expected Colour-Magnitude for elliptical
galaxies formed at $z_{\rm for}$=10.
}}
\label{fig:cmex1}
\end{center}
\end{figure*}

\addtocounter{figure}{-1}

\begin{figure*}
\begin{center}
\epsfxsize=16cm
\begin{minipage}{\epsfxsize}{\epsffile{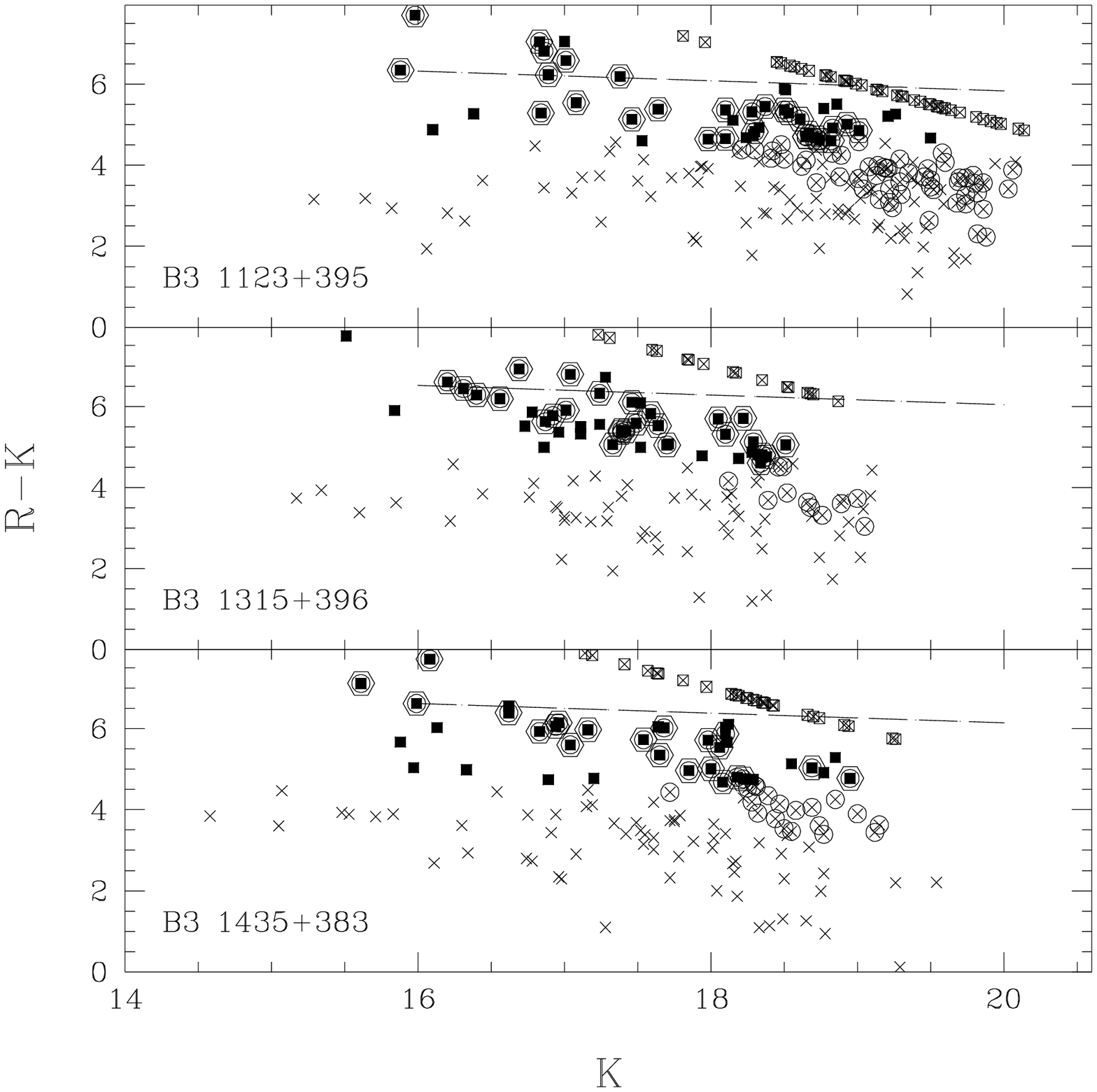}}
\end{minipage}
\caption[]{{\small
Continued
}}
\label{fig:cmex2}
\end{center}
\end{figure*}

\addtocounter{figure}{-1}

\begin{figure*}
\begin{center}
\epsfxsize=16cm
{\epsffile{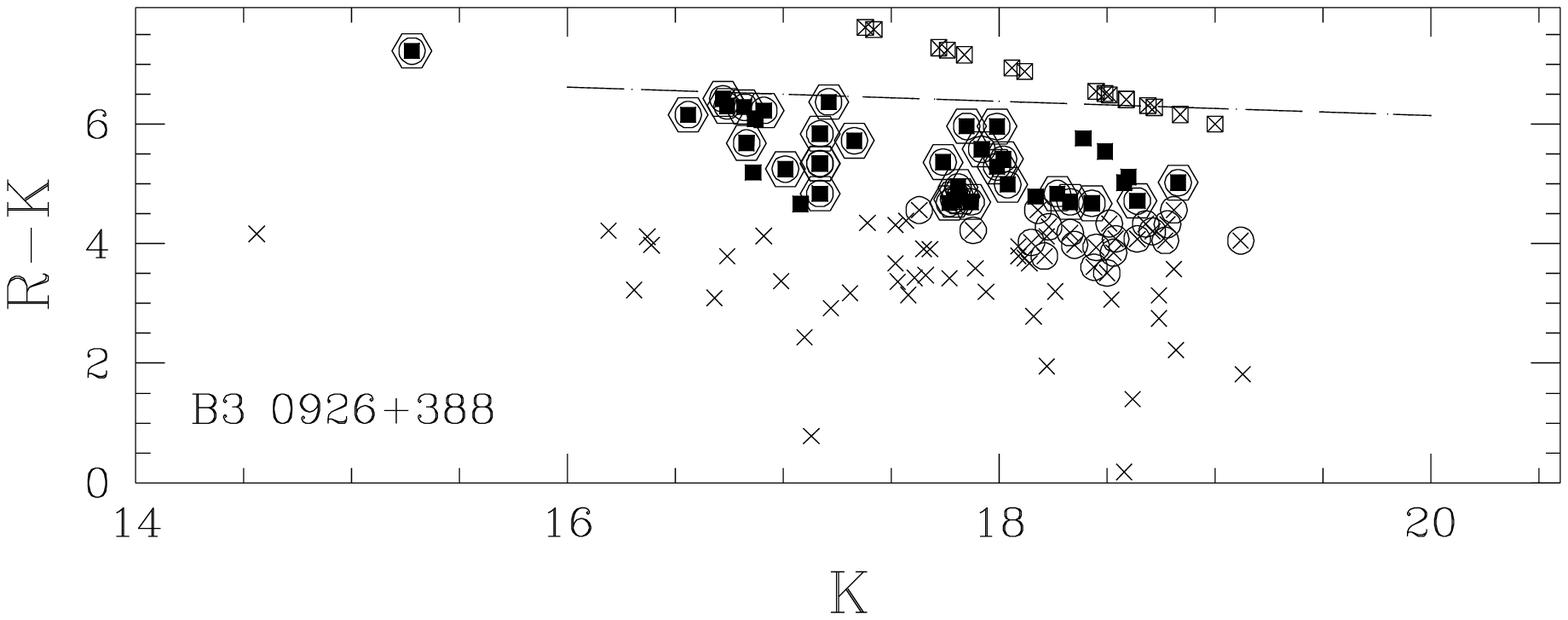}}
\vspace{-10cm}
\caption[]{{\small
Continued
}}
\label{fig:cmex3}
\end{center}
\end{figure*}

Figure \ref{fig:cmex1} shows the distribution of the $R-K$ colours
versus the $K$-band magnitude for the galaxies detected in each
field. The discontinous line shows the expected CM relation for a bulk
of galaxies formed at $z_{\rm for}\sim$10, assuming a passive evolution
(Bruzual-Charlot $c$-model, $\tau$=1Gyr, $z_{\rm for}$=10, initial
mass function from Salpeter between 0.1 and 125 M$_{\odot}$), with a
constant slope. The $z_{\rm for}$ was selected to match the
zero-points of the CM relation of a sample of spectrocopically
confirmed clusters, with a range of redshifts between $z\sim$0.2-1.3
(Kodoma et al. 1998). Open squares show objects detected in $K$ but
not in $R$ band ($R-K$ lower limit). Solid squares show the galaxies
with $R-K$ colours less than 1.5 mag from the expected CM relation,
i.e., candidates to be cluster members. Open circles show faint
galaxies with red colours in the optical bands ($B$$>$23.5, $R$$>$22,
$B-R$$>$1.5), and from them, open hexagons show the galaxies with red
optical-NIR colours ($R-K$$>$4.5). As we discussed above, the CM
relation should be produced by this last group.

It is seen in Figure \ref{fig:cmex1} that open circles conform an
upper envelope in the colour-magnitude diagram, which traces the the
expected CM relation. The hexagons (old galaxies, without recent star
formation) reproduce this relation more accurately, although the
dispersion is larger than found in lower redshift samples
(Arag\'on-Salamanca 1993; Stanford et al. 1997). The fact that it is
possible to reconstruct the red sequence support our adopted
procedure to select a sample of cluster candidates.

The relation shows a larger dispersion for the fields around the
quasars with highest redshifts (Fig. \ref{fig:cmex1}). This increase
of the dispersion could be due to three main causes: (a) an increase
of the contamination from field galaxies, extimated to be as large as
a $\sim$20\%, (b) an increase of the photometric errors of $R$ and $K$
magnitudes, stronger for the faintest galaxies, which ranges between
0.14--0.42 magnitudes, (c) an intrinsic increase of the dispersion.
It is clear that our data are affected by (a) and (b), which could
produce the observed increase. However, (c) has to be taken in account
as well: (i) if the CM relation is due to bulk formation of the
cluster population, an increase of the dispersion towards higher
redshifts would be expected (if we assume a range of
$z_{\rm for}\sim$2.5--5, at $z$=1.6 we are at $\sim$0.4--1 Gyr from the
formation time), (ii) in case of latter starformation process (Tanaka
et al. 2000; Haines et al. 2001), an increase of the dispersion is
also expected. It is needed to reduce at maximum the effects of (a)
and (b) to determine the quantative contribution of (c) to the
observed increase of the dispersion.

\section{Properties of cluster candidates}

\subsection{$BRK$ colour distribution}

The number of galaxies selected to be cluster candidates using the
above described procedure ranges between 40 and 47 for each
field. This number is similar to the number of galaxies that produces
the excess field-to-field, which ranges between 30 and 40.
Furthermore, the differences between both numbers is $\sim$20\%,
which is the extimated fraction of misclassified galaxies.

Only a fraction between $\sim$15\% and $\sim$41\% of the galaxies
selected from their $R-K$ colours were detected in the $B$ band. The
$B$-band images were less deep than the $R$-band, therefore only
the bluer galaxies with $B-R\la$1.5 mag of cluster candidates
were detected in this band. These galaxies form a population of red
galaxies in the optical-NIR range ($R-K$$\ga$4.5) and blue in the
ultraviolet range ($B-R<$1.5, $\lambda\sim$1760\AA-2200\AA\ at
$z\sim$1.5). This population has been found before in other clusters
at similar redshifts (Tanaka et al. 2000; Haines et al. 2001).

\begin{figure*}
\centering
\begin{center}
\epsfxsize=16cm
\begin{minipage}{\epsfxsize}{\epsffile{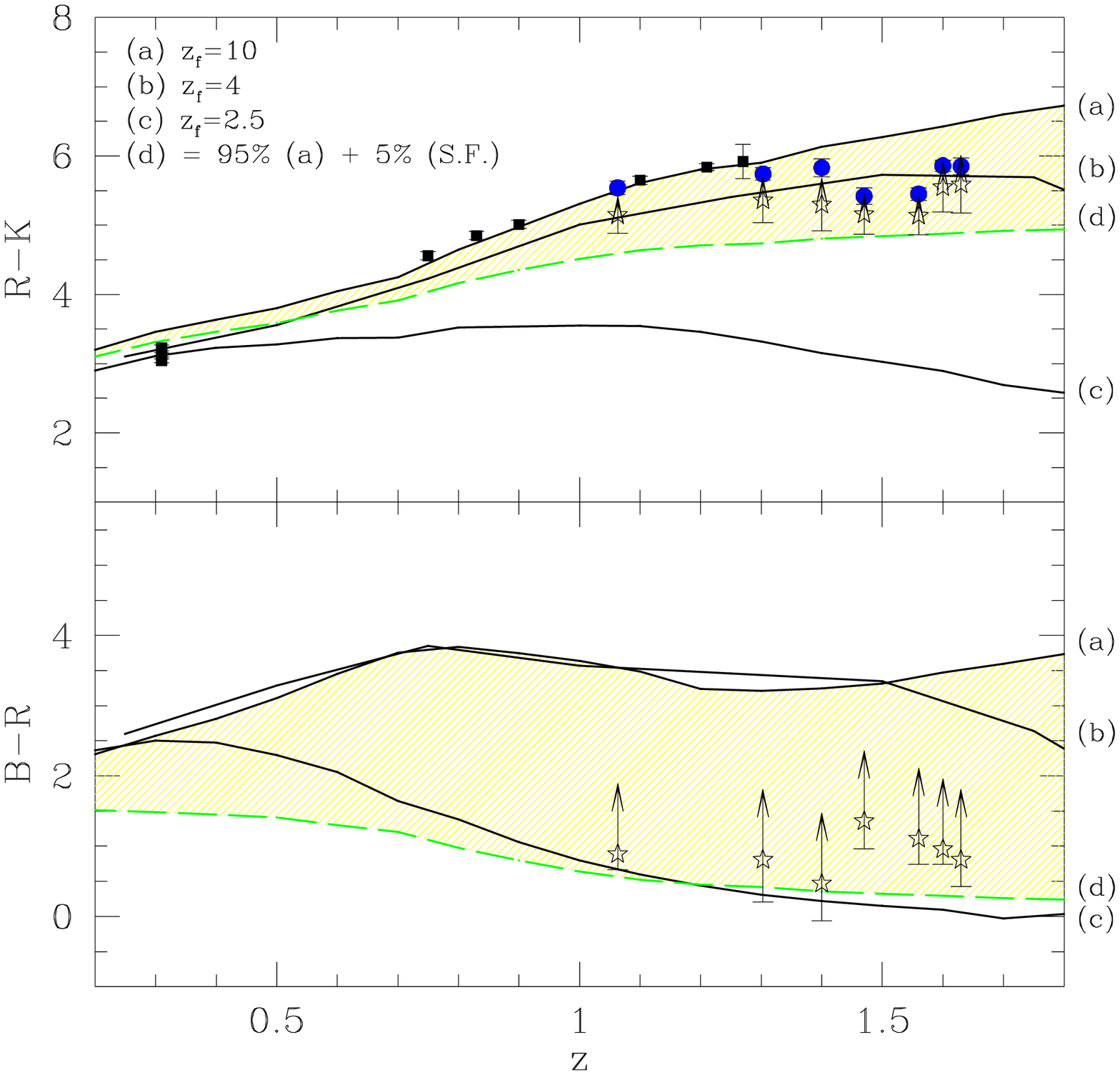}}
\end{minipage}
\caption[]{{\small
$R-K$ and $B-R$ colour distribution along the redshift
for two galaxies evolution model: (i) passive evolution (back lines) 
with different formation redshifts ($z_{\rm for}$=2.5,4,10) and (ii)
passive evolution plus a fraction of galaxy mass with a recent
starformation (green line). The yellow shaded region show the second
model for different fractions of mass under starformation. 
Black squares show the zero-points of the CM correlation for
spectroscopic confirmed clusters with redshifts between $z\sim$0.2-1.3.
Blue circles show the zero-points found for our clusters, and stars
show the mean colours for the galaxies candidates to cluster.
}}
\label{fig:CM}
\end{center}
\end{figure*}

Figure \ref{fig:CM} shows the $R-K$ and $B-R$ colour distribution
versus the redshift for two different evolution models: (i) passive
evolution, with three different formation redshifts ($z_{\rm
for}$=2.5,4 and 10), black lines, and (ii) a second model in which the
bulk of the galaxy follows passive evolution (the previous model, with
$z_{\rm for}$=10), but with a fraction of the mass showing more recent
star formation. The dashed line shows the result for a fraction of
$\sim$5\% under star formation, while the shaded region traces the
evolution when this fraction reduces from 5\% to 0\% (i.e., passive
evolution). Solid squares show the zero points of the CM relation for
spectroscopically confirmed clusters with a redshift range between
$z$=0.3 and 1.3 (Kodoma et al. 1998). It is seen that passive
evolution models with high formation redshifts ($z_{\rm for}>$4),
describe the observational data accurately. The zero points of CM
relation from our cluster candidates are shown as solid circles. As
they were selected based on this CM relation, it is expected that they
fit quite well with a passive evolution model for old galaxies.

We have plotted the mean colours of candidate galaxies as open stars.
The $R-K$ colours present an evolution compatible with old elliptical
galaxies, whereas the $B-R$ are clearly bluer. It should be noted here
that the $B-R$ colours are a lower limit, since we have detected only
a fraction of the candidates in the $B$-band images. Therefore, the
range of $B-R$ colour is expected to be wider, from plotted data to
values expected for old elliptical galaxies. A first conclusion that
could be extracted from Figure \ref{fig:CM} is that the observed
distribution rejects the possibility of a young population formed at
low redshifts ($z_{\rm for}\la$3). Although the $B-R$ colours of some
galaxies ($\sim$25\%) were compatible with this hypothesis, their
$R-K$ colours are clearly redder than expected.

A pure passive evolution model, with galaxies formed at $z_{\rm
for}>$2.5, cannot explain the data either. Although the $R-K$ colours
fit accurately to this model, { the observer range of $B-R$ colours
from those galaxies detected at $B$ does not agree with it}. Only a
mixed model, where clusters present recent star formation, could
explain the observed data. With a reduced fraction of the galaxy mass
under starformation ($\sim$5\%), juxtaposed to passively evolving
population, it is possible to describe the observed data. It is needed
to note here that the mixed model does not mean that $\sim$5\% of the
cluster galaxies were formed at lower redshifts. In that case, the
dispersion of the $R-K$ colours would be larger than observed. What we
found is that clusters contain a population of old elliptical
galaxies, and a fraction of them ($\sim$25\%) has undergone recent
star formation including $\la$5\% of their masses.

Haines et al. (2001) and Tanaka et al. (2000) presented similar
results for two clusters at $z\sim$1.3 and $z\sim$1.1,
respectively. They found a larger fraction ($\sim$50\%) of red
galaxies with ultraviolet excess compared with $\sim$10-40\% on our
clusters, but their physical interpretation was similar. Our
estimation of the fraction of {\it blue} galaxies is similar to those
found in clusters at $z\sim$0.9 (Postman et al. 1998a; Lubin et
al. 1998; Racos \& Schombert 1998), but larger than the fraction of
these objects found at lower redshifts, $z\sim$0.2 (Smail et
al. 1998). Spectroscopic studies of clusters at $z\sim$0.3--0.5, have
shown that there is a significant fraction of galaxies in post star
formation periods (e.g., Dressler \& Gunn 1992; Couch et al. 1994,
1998; Poggianti et al. 1999). It seems that these galaxies has
undergone star formation at higher redshifts, and they could be the
fossils of the galaxies with ultraviolet excess found in higher $z$
clusters.

\subsection{Spatial distribution}

\begin{figure*}
\centering
\begin{center}
\epsfxsize=8cm
\begin{minipage}{\epsfxsize}{\epsffile{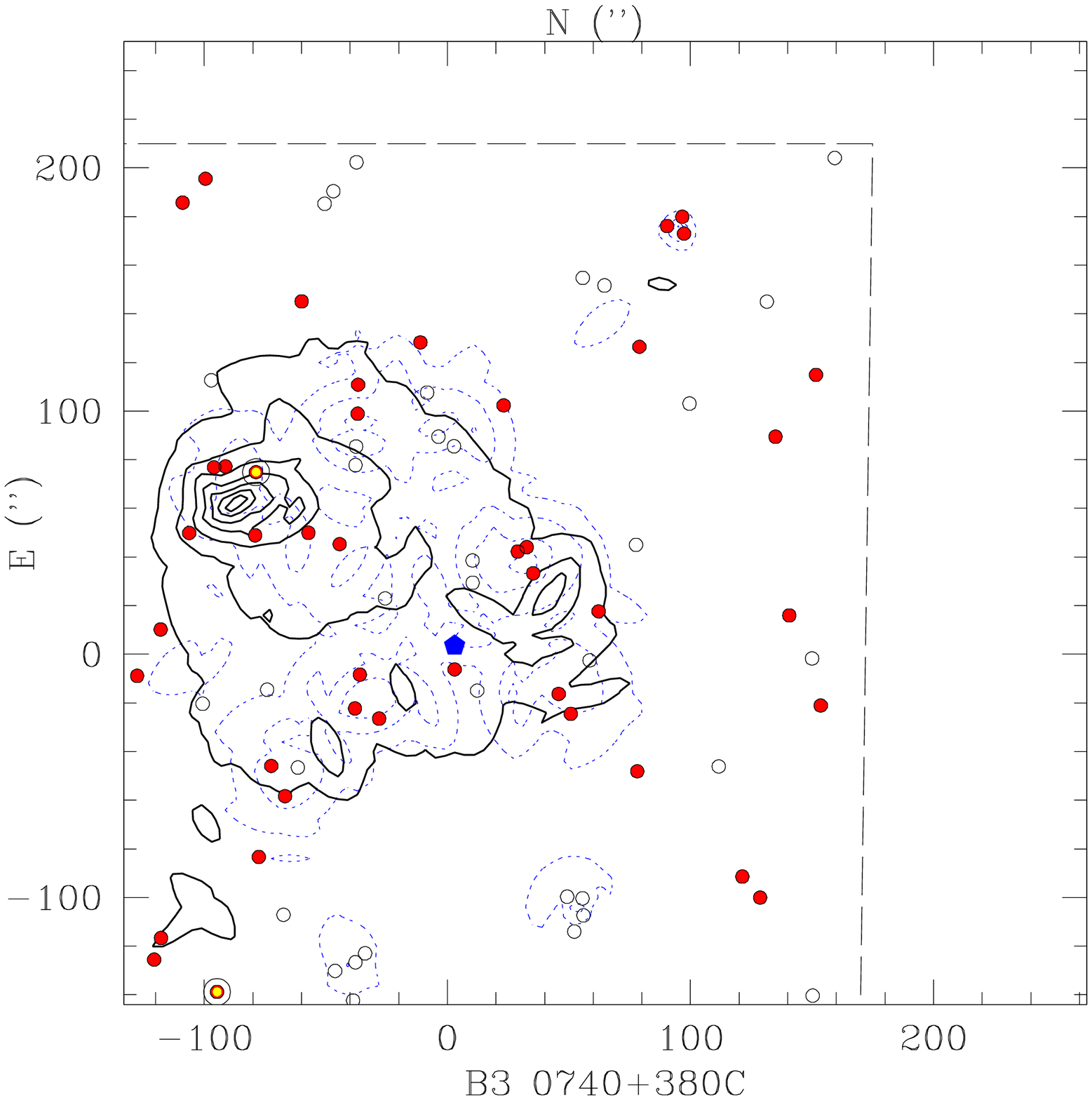}}
\end{minipage}
\epsfxsize=8cm
\begin{minipage}{\epsfxsize}{\epsffile{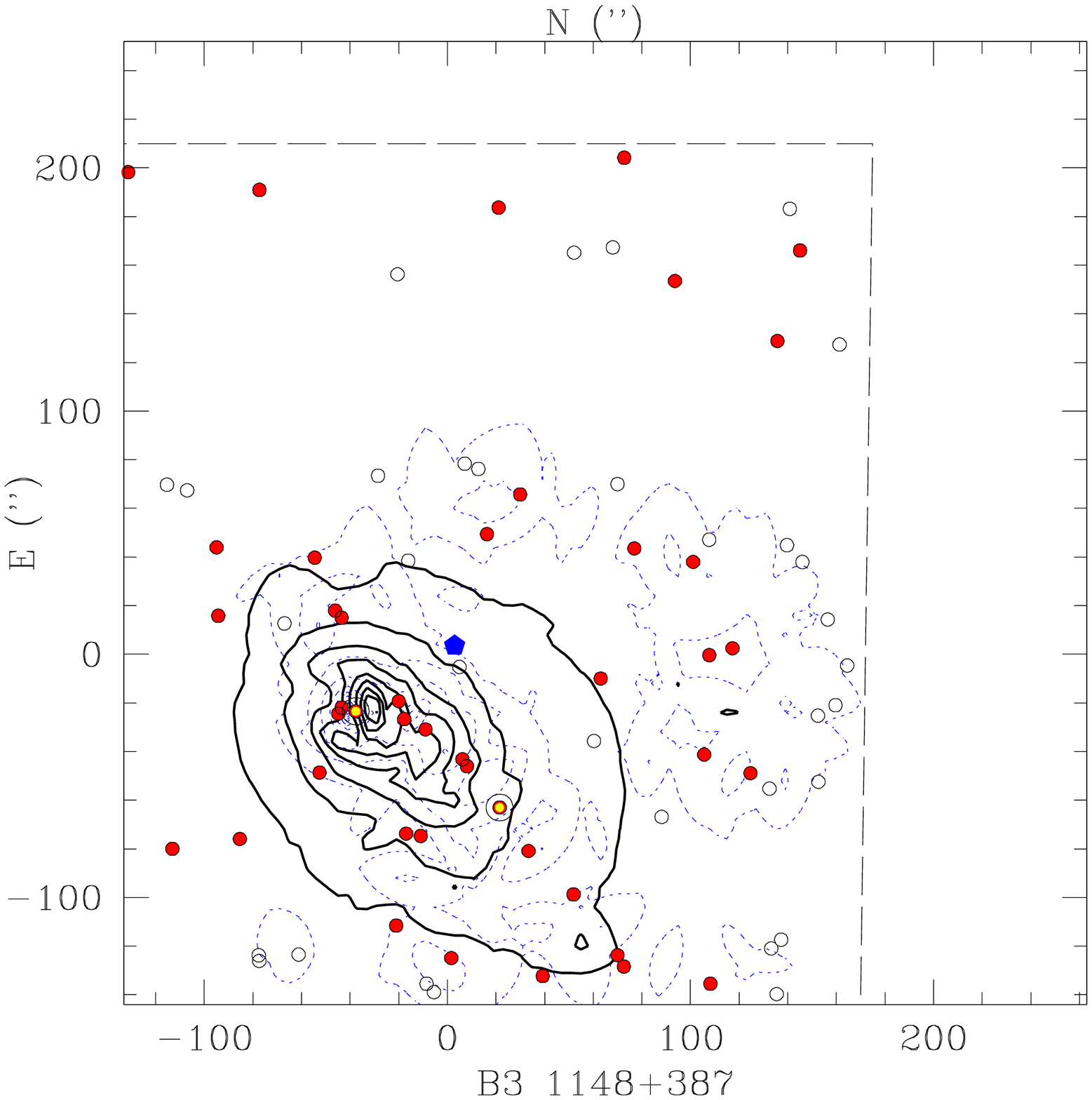}}
\end{minipage}
\epsfxsize=8cm
\begin{minipage}{\epsfxsize}{\epsffile{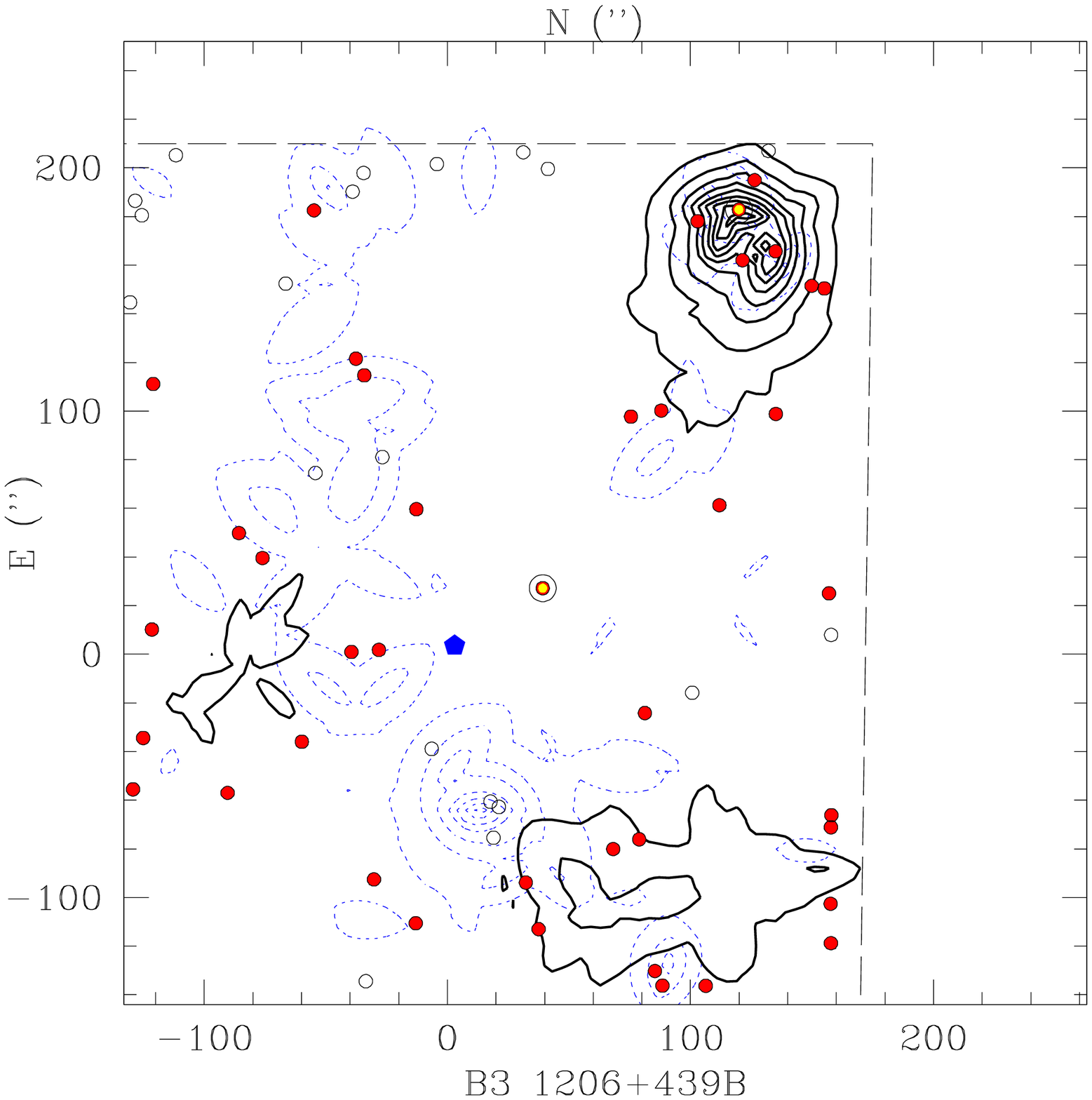}}
\end{minipage}
\epsfxsize=8cm
\begin{minipage}{\epsfxsize}{\epsffile{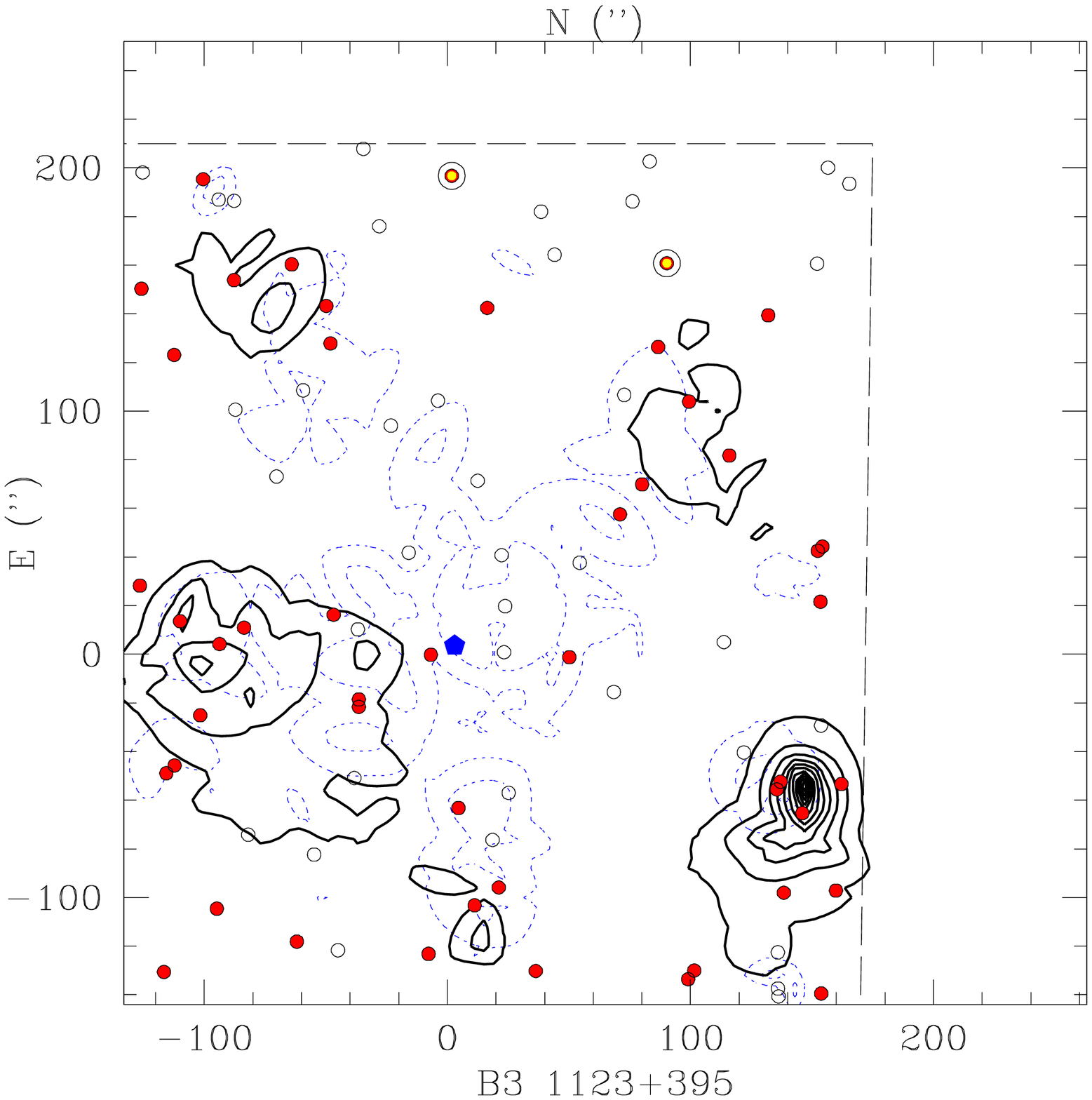}}
\end{minipage}
\end{center}

\caption[ ] {{\small Spatial distribution of the candidates cluster
members around the quasars in the overlapping region of the optical
and NIR images (limited by dashed lines). Red solid circles show the
galaxies selected as cluster members, whereas open circles show
galaxies detected in $K$ but not in $R$, ie, red enough to be cluster
members.  Black contours show the density estimation of the cluster
candidates (red solid circles), whereas blue dotted contours show this
value for all the plotted galaxies. First contour is at 1$\sigma$ over
the mean value, being this value the distance between sucesive contour
levels. Quasar position is indicated as a pentagone for each
field. Yellow solid circles show the position of the two brighest
cluster galaxies. 
 }}
\label{fig:dmap1}
\end{figure*}

Figure \ref{fig:dmap1} shows, for each field, the bidimensional
spatial distribution of the galaxies selected as cluster candidates
(solid circles), together with galaxies detected in $K$ but not in $R$
band (open circles). A fraction of this latter group could be also
members of the clusters (Figure \ref{fig:cmex1}). We have also plotted
in Figure \ref{fig:dmap1} an estimation of the cluster density,
determined using the variable kernel method (used in Paper I;
Silverman 1986) with the Epanechnikov kernel (Epanechnikov 1969). This
method is an improvement of the nearest neighbour method, matching the
smoothing factor to the local density for each point.  The density
estimation generally used is based in the number of objects counted in
a certain box of fix size, centered in the interest point. However,
the estimator used here is inversely proportional to the area needed
to contain a certain number of observables, $k$ (galaxies in our
case), and inversely proportional to a smoothing parameter $h$. The
number of observables $k$ was fixed to mean galaxy density for each
image. This density estimation reflects the properties of the
distribution better than the more extended used procedure (read
discussion on Silverman 1986).

The density estimation of candidate galaxies is shown as a continous
contour, whereas the density estimation for all the galaxies plotted
in figure (solid circles and open circles) is shown as a dotted
contour. The quasar position is indicated with a pentagone, and the
two brighest galaxies in each field with two concentric open and solid
circles. Figures are centered on the field-of-view of 
$K$-band images, showing the dashed lines the limit of the overlapping 
region with optical images (i.e., the region sampled by $B$, $R$ and $K$ band
images).

As we found in Paper I for the overdensities of galaxies in the
optical images, the quasars are not located in the center of the
clusters, but at $\sim$100$\arcsec$ far from them ($\sim$1Mpc at the
mean redshift of the quasars). This off-centering is observed for the
distribution of the cluster candidates, even including the galaxies
not detected in the $R$ band. The peaks of both samples match only for
the two quasars at lowest redshift, for which the density shows a
3$\sigma$ unique peak, being similar to the spatial distribution of
well known low-$z$ clusters. For the remaining fields at higher
redshift, the distribution is more similar to groups of galaxies or
proto-clusters than totally assembled clusters. All of them shows a
peak up to 6$\sigma$ that can be considered as the center of the
cluster. The radial profiles of the density around these peaks fit to
a King model, with core radii ranging between 202
kpc$\la$r$_c$$\la$850 kpc, and a mean value of r$_c$$\sim$570
kpc. This result reinforces our suspicion that we have found clusters
(or proto-clusters) at the redshift of the quasars.

The fact that quasars do not inhabit the more dense regions of the
clusters (core) agrees with the hypothesis of a merging process as the
origin/feeding mechanism of the nuclear activity. Aarseth \& Fall
(1980) demonstrated that for merging process to work, velocities at
encounters between galaxies should not exceed significantly their
internal velocity ($\sim$200 km s$^{-1}$). Similarly, Bekki \& Noguchi
(1994) demonstrated using simulations that the upper limit to the
relative velocity of the gas clouds of galaxies that collide should be
$\la$600 km s$^{-1}$ to allow these clouds to concentrate within 10
kpc, once the collision takes place. This is a first condition to
allow these clouds to feed the AGN. This implies that in cluster cores
(where velocity of galaxies is $\gtrsim$600 km s$^{-1}$) the merging
processes due to collisions were less efficient than in the outer
regions of the clusters, where the velocity of galaxies
decreases. Similar results have been found by other authors at similar
redshift ranges (Haines et al. 2001; Tanaka et al. 2000; Thompson et
al. 2000; Nakata et al. 2001).  The differences in populations of
galaxies at the core and outer regions of clusters (with a larger
fraction of gas rich galaxies) could play a role in this process.

In Figure \ref{fig:dmap1} we have indicated the position of the two
brightest cluster galaxies (BGC) as concentric open and solid circles
(using $K$-band magnitudes). In four of the seven fields one of these
two galaxies inhabits regions close to the core of the cluster, three
of them are the lowest redshift objects. Similar features have been
found in well known low-$z$ clusters, where it is usual to find a cD
galaxy in the cluster core. We have not found this effect on the
fields at highest redshifts, even in the case that we take in account
the galaxies not detected in the $R$ band. This result, together with
the lack of uniformity in their spatial distribution, could indicate
an evolution of structure. However, a spectroscopic confirmation of
the cluster candidates, and a study over larger, statistically
significant samples is needed before a firm conclusion can be made.

\addtocounter{figure}{-1}

\begin{figure*}
\centering
\begin{center}
\epsfxsize=8cm
\begin{minipage}{\epsfxsize}{\epsffile{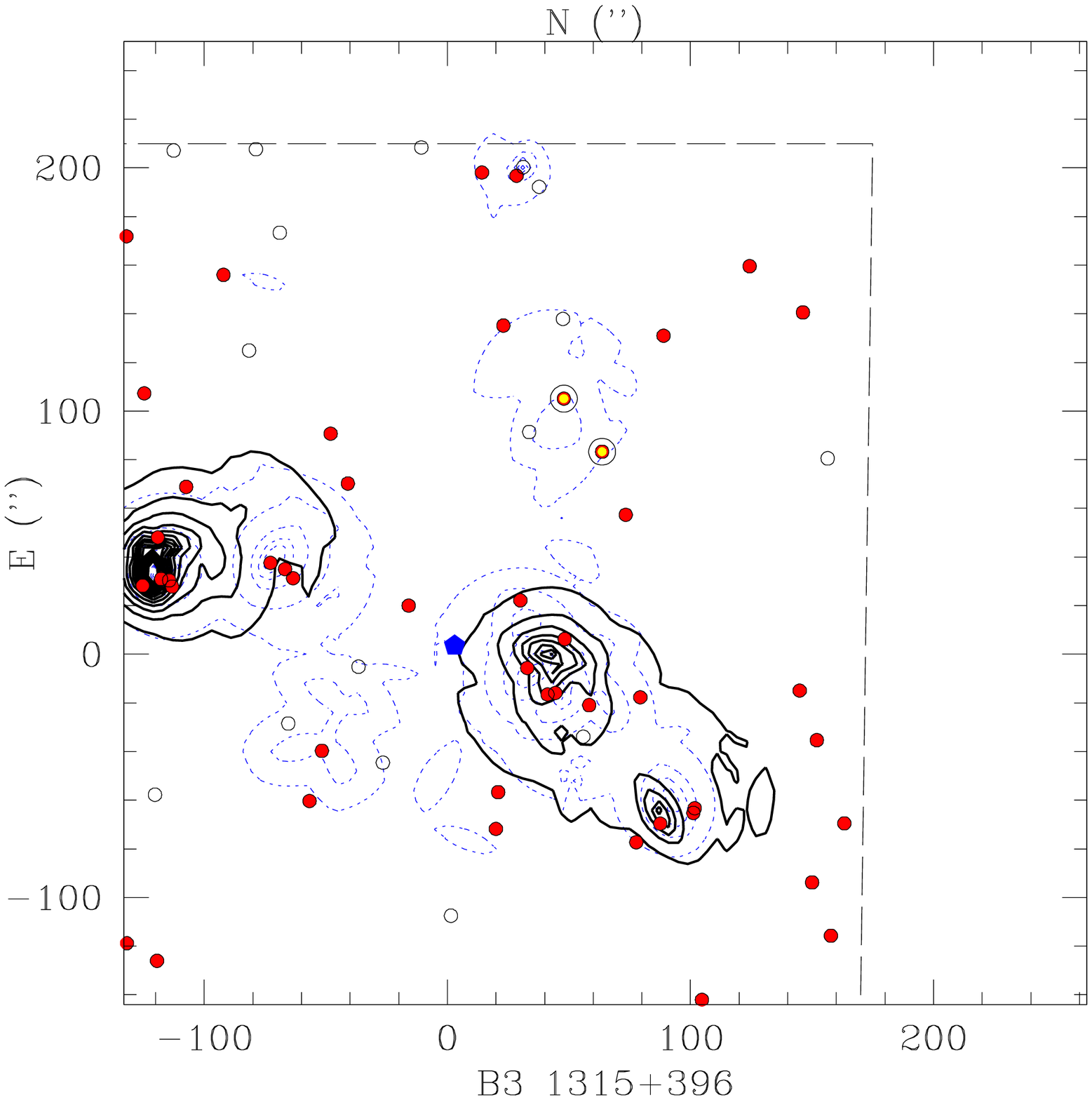}}
\end{minipage}
\epsfxsize=8cm
\begin{minipage}{\epsfxsize}{\epsffile{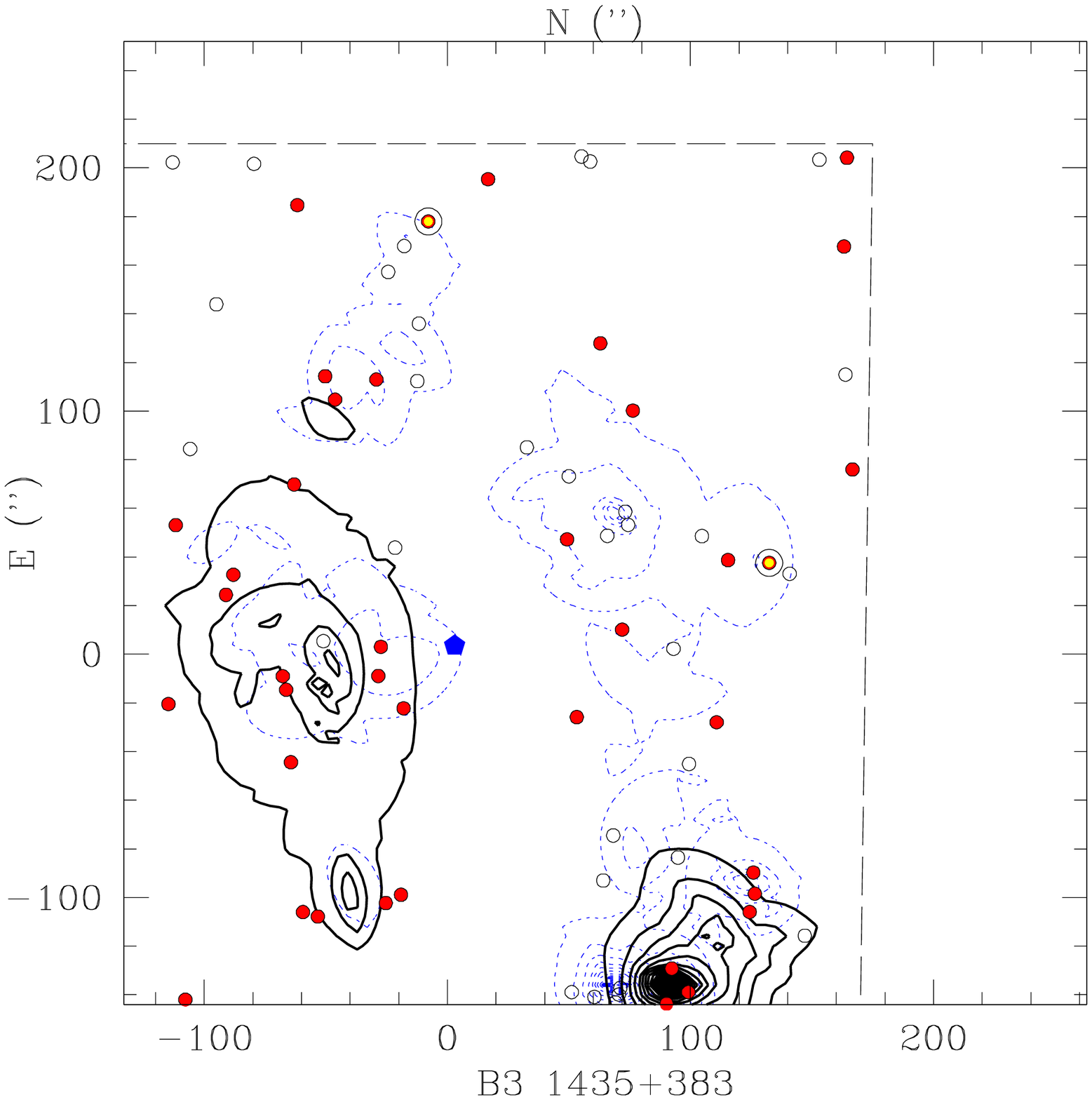}}
\end{minipage}
\epsfxsize=8cm
\begin{minipage}{\epsfxsize}{\epsffile{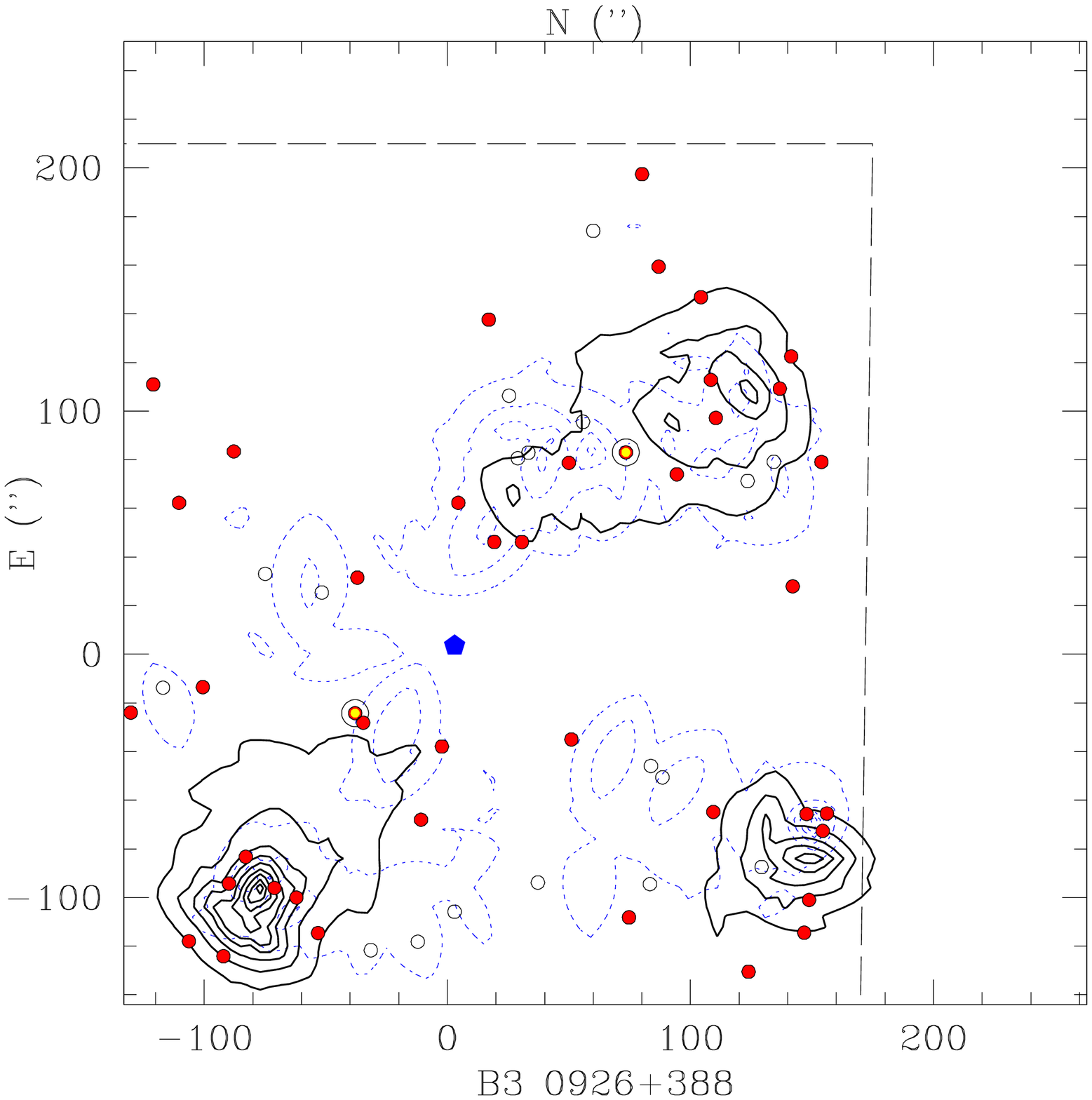}}
\end{minipage}
\end{center}

\caption[]{{\small
Continue
}}
\label{fig:dmap2}
\end{figure*}

\section{Discussion and Conclusions}

In this article we have analyzed deep $K$-band images around seven
radio loud quasars, with a redshift range 1$\la$$z$$\la$1.6.  In Paper
I, we presented the analysis over deep $B$ and $R$ band images around
the same quasars. Using the present results together with the results
from Paper I, we found that there is a significant excess of galaxies
around these quasars, with magnitudes, colours, spatial scale and
number of galaxies compatible with clusters of galaxies at the
redshift of the quasars. This result agrees with previous studies of
possible overdensities of galaxies around radio-loud quasars (Hintzen
et al. 1991; Hall \& Green 1998), in contrast with results obtained on
similar studies over radio-quiet samples (Boyle \& Couch 1993; Teplitz
et al. 1999), although some of these studies found significant
excesses around radio-quiet quasars (Haines et al. 2001). These
results also agree with studies based on samples of lower redshift
quasars, where the environment of radio-loud and radio-quiet quasars
seem to present significant differences (Yee
\& Green 1987; Elligson et al. 1991).

There are different mechanisms that could explain the described
differences: (i) it is expected that radio loud sources inhabit gas
rich clusters, if the intracluster gas plays a substancial role as the
fuel of the radio source. It is reasonable to assume that intracluster
gas increases with cluster density, which could explain the
differences in the environments of radio-loud and radio-quiet sources;
(ii) extended radio emission, with steep-spectrum emission, is
normally explained as the result of the interaction of the jets with
the intergalactic medium (e.g. Miley 1980). This medium confines the
radio emission lobes. It is expected that the intergalactic medium was
denser in more populated clusters, and therefore, there could be a
relation between cluster density and radio loudness. Maybe the
combination of both mechanism is needed to explain the observed
differences. It is interesting to note here that the second mechanism
could explain the correlation between the excess of galaxies and the
slope of radio emission found by Hall \& Green (1998) and Mendes de
Oliveira et al. (1998).

Two mechanisms have been proposed to explain the connection between
nuclear activity and its environment: the dynamic state of the
clusters and the evolution of intracluster gas (Yee \& Ellingson
1993). In both schemes the merging/interaction processes between the
host galaxy and its neighbours play a substancial role, which
reinforce the hypothesis that a merging process could be the origin of
nuclear activity (Barnes \& Hernquist 1992; Shlosman 1994; Bekki
1999). There is ample evidence to support this idea, like the large
fraction of host galaxies under possible interaction/merging
processes, their large luminosities, and possible traces in their
spectra (Hutchings 1987; Disney et al. 1995; Hutchings \& Neff 1997;
Carballo et al. 1998; S\'anchez et al 2002; Nolan et al. 2001).

We found an envelope of red galaxies in the colour-magnitude diagram
($R-K$ vs. $K$), using the combined information from the three
bands. This envelope was consistent with the CM relation found for
clusters with a redshift range between 0.3$<$$z$$<$1.3. Extrapolating
this relation to the redshift range of our sample (interpolating in
some cases) we selected a sample of cluster candidates. Among these
galaxies selected as candidates by their $R$ and $K$ magnitudes, only
$\sim$25\% were detected in the $B$ band. This group of galaxies
comprises a subsample of red galaxies with ultraviolet excess, similar
to found by Tanaka et al. (2000) and Haines et al. (2001).

The colour evolution of the cluster candidates was compatible with a
model in which the bulk of the galaxies follows a passive evolution
typical of old galaxies ($z_{\rm for}\sim$4-10), whereas a fraction of
$\sim$25\% follows a mixed evolution. This mixed evolution implies
that at least 95\% of the mass follows a passive evolution and a
fraction lower than $\sim$5\% has undergone presents star formation,
even at the observation redshift. This star formation could be induced
by collision processes (Lubin et al. 1998; Tanaka et al. 2000; Haines
et al. 2001). Spectroscopic studies of these galaxies are needed to
clarify these results.

We have found that the quasars of our sample do not inhabit the
cluster cores. The position of the quasars in the clusters is also
compatible with the hypothesis of a merging process as the QSO origin.
Only collisions between galaxies with relative velocity similar to
their internal velocity could induce merging. This merging was a
prerequisite for the infall of mass into the inner region and the
formation or feeding of the nuclear source. This limits the region
where collisions could lead to merging to the outer region of the
clusters, since the velocity were too large in the cluster cores.

Summarizing, out results suggest that radio-loud quasars inhabit the
outer regions of low populated clusters at high redshift. Clusters
that show mixed evolution, which main component would be passive
evolution for old galaxies with a reduced fraction under a recent
starformation process. These results support the hypothesis of a
merging process as the origin/feeding mechanism of nuclear activity.




\begin{acknowledgements}

We thank all Calar Alto observatory staff for the friendly
support and the unestimable help during the observing runs.
S.F.S\'anchez thanks Danny Lennon (ING Head of the astronomy group)
for its support his the realization of this research.

We thank L. Wisotzki for his kind and unestimable help in the
correction of the poor english of the first version of this 
article, and for his useful comments.

We thank Dr. Stockton, who refereed this article, for valuable
comments, which help to improve the present article.

S.F.S\'anchez would like to thank Eresmas S.A.,  and in particular
to Mariano Tejedor, for their support with time and computers to
finish this article.

J.I. Gonzalez-Serrano wants to aknowledge financial support from
DGESIC (Ministerio de Educacion y Cultura) under project
PB98-0409-C02-02.

\end{acknowledgements}



\end{document}